\documentclass[%
 reprint,
 superscriptaddress,
 amsmath,amssymb,
 prl
]{revtex4-2}
\usepackage{graphicx}% Include figure files
\usepackage{dcolumn}% Align table columns on decimal point
\usepackage{bm}% bold math
\usepackage{csquotes}
\usepackage[english]{babel}
\usepackage{siunitx}
\usepackage{booktabs}
\usepackage{tabularx}

\usepackage{braket}
\usepackage[normalem]{ulem}
\usepackage[caption=false, justification=centerlast]{subfig}
\usepackage{todonotes}

\newcommand{\rom}[1]{\uppercase\expandafter{\romannumeral #1\relax}}

\usepackage[%
colorlinks=true,
urlcolor=blue,
linkcolor=blue,
citecolor=blue
]{hyperref}

\usepackage{tabularx}

\begin{document}
	      
%\preprint{APS/123-QED}

\title{Phase-dependent bubble hosing and resonant amplification of betatron oscillation in few-cycle laser wakefield accelerator}% Force line breaks with \\

\author{Bifeng Lei}
\email{blei@pku.edu.cn}
\affiliation{Center for Applied Physics and Technology, HEDPS, and SKLNPT, School of Physics, Peking University, Beijing 100871, China}
\affiliation{Department of Physics, The University of Liverpool, Liverpool, L69 3BX, United Kingdom}

\author{Daniel Seipt}
\affiliation{Institute of Optics and Quantum Electronics, Max-Wien-Platz 1, 07743 Jena, Germany}
\affiliation{Helmholtz-Institute Jena, Fröbelstieg 3, 07743 Jena, Germany}

\author{Bin Liu}
\email{liubin@glapa.cn}
\affiliation{Guangdong Institute of Laser Plasma Accelerator Technology, Guangzhou, China}

\author{Matt Zepf}
\affiliation{Institute of Optics and Quantum Electronics, Max-Wien-Platz 1, 07743 Jena, Germany}
\affiliation{Helmholtz-Institute Jena, Fröbelstieg 3, 07743 Jena, Germany}

\author{Carsten Welsh}
\affiliation{Department of Physics, The University of Liverpool, Liverpool, L69 3BX, United Kingdom}

\author{Bin Qiao}
\email{bqiao@pku.edu.cn}
\affiliation{Center for Applied Physics and Technology, HEDPS, and SKLNPT, School of Physics, Peking University, Beijing 100871, China}
\affiliation{Frontiers Science Center for Nano-optoelectronic, Peking University, Beijing 100094, China}

\date{\today}% It is always \today, today,

\begin{abstract}
%%%
Intrinsic hosing instability of the few-cycle laser-driven plasma bubble in transverse directions has been theoretically investigated by the perturbation theory and confirmed by particle-in-cell simulations.
It is caused by the residual momentum of the temporally asymmetric laser-plasma interaction.
The electron beam trapped in such a bubble undergoes the amplified betatron oscillation as a result of the linear and parametric resonance.
By identifying the resonance conditions and thresholds, these amplification processes can offer a novel way of manipulating the betatron dynamics to fullfil the needs of various advanced applications. 
\end{abstract}

\maketitle

%\tableofcontents
%\section{Introduction}

%\begin{itemize}
%	\item Asymmetry: spatial and temporal
%	\item Chirp 
%\end{itemize}

The development of laser wakefield accelerator (LWFA) has attracted much attention because it holds the promise of providing low-cost, high-energy electron beams with far-reaching impact on many areas of nature science, such as in strong field quantum electrodynamics{~\cite{Poder2018EXP,Cole2018EXP, turner2022strong}}, free electron laser~\cite{Wang2021FREE, labat2022seeded,pompili2022free,Galletti2022stable}  and particle physics~\cite{Schroeder2010,SCHROEDER2016113,Tanaka2020cur}.
Over the past two decades, great progress has been seen in LWFA, such as electron beams with energy close to 8~\si{GeV} accelerated in a 20 centimeter gaseous plasma~\cite{Gonsalves2019PW}.
In such experiments, the high accelerating gradient, in order of $100~\si{GV/m}$, is generated by the ponderomotive force~\cite{kruer2019physics}, where the quiver motion of electrons in the laser field does not couple with the plasma dynamics on the scale of plasma wavelength. As a result, the details of the electron motion on the scale of laser wavelength can be removed by averaging over time~\cite{Mora1997kin,Esarey2009PHY}. This approximation is the so-called time-averaged ponderomotive approximation, or laser envelope approximation~\cite{Shukla1986aa}.

However, the ponderomotive approximation does not provide complete accuracy in describing the high-intensity ultrashort laser-plasma interaction, for example, if plasma density gradients are sharp or shape of the laser pulse changes rapidly~\cite{Terzani2021acc,Tuev:2023aa}. 
This has been observed in some experiments and simulations where the realistic laser pulses of the spatial and temporal asymmetries are used~\cite{Ferri2016aa, Corde2013aa}. 
%This is true for few-cycle or temporally asymmetric pulses, where the plasma electrons transmit the laser pulse in a short amount of time. 
%%
In these cases, averaging over interaction time fails to remove the quiver motion, which can then be cumulated to give the residual momentum of the scattered plasma electrons.
As demonstrated by Nerush et al.~\cite{Nerush2009carr}, the residual momentum is dependent on the carrier-envelope phase (CEP) of the laser pulse and can be significant for a few-cycle or temporally asymmetric laser pulse. It may result in an intrinsic bubble hosing instability in the direction of laser polarization direction~\cite{Seidel2022pointing}.
This CEP effect has high impact on laser-plasma interaction~\cite{Pollock2015for, Ma2016aa, Satanin2014amp, Ferri2016aa} and has been explored for various applications, including the control of electron injection~\cite{Kim2021sub, Huijts2022wave}, stabilisation of beam jitter~\cite{Seidel2022pointing}, and optimisation of betatron radiation~\cite{Chen2021enh, Rakowski2022aa, Mishra2022enh}.

%The resonant amplification of betatron oscillation is a widely considered way to enhance the betatron radiation, which is driven by the coupling between the laser and the transverse focusing field. It usually occurs in a stable bubble where neither centroid nor sheath oscillation exist, and in the dephasing region where the trapped electron beam catches up the tail of the lase pulse~\cite{Cipiccia2011ga, Nemeth2008laser, Curcio2015re}. The recent experiment has shown an significant enhancement by using an additional copropagating laser pulse to modulate the trapped electron beam from beginning~\cite{Horny2020att}.

In this paper we present a different type of the bubble hosing instability and its corresponding resonant amplification of the betatron oscillation in the few-cycle or temporally asymmetric laser-driven LWFA. 
This hosing effect is intrinsic in both transverse directions and expressed in the form of the bubble centroid oscillation (BCO) and the bubble sheath oscillation (BSO). 
BCO is caused by the residual momentum from the laser-plasma interaction and is stronger in the laser polarization direction. It can result in the linear resonance which can amplify the betatron oscillation if the betatron frequency is close to the centroid oscillation frequency. 
BSO is caused by the fast oscillation of ponderomotive force and can deform the bubble sheath. 
BSO can result in the additional amplification of betatron oscillation through parametric resonance by coupling with the potent BCO effect.
By giving the analytical amplification conditions and the parameter regime where the parametric amplifications become important, our findings potentially offer a mechanism to actively manipulate the betatron dynamics in the plasma bubble for a range of advanced applications.
%%%%%%%%%%%%%%%%%%%%%%%%%%%%%%%%%%%%%

%\section{Bubble hosing}
Relativistic laser plasma interaction can be studied by using the perturbation theory~\cite{Nerush2009carr}. 
We firstly start from a single electron moving inside a laser field propagating along $z$ axis. The equation of motion can be written as
\begin{align}
	\frac{d^2 \bm{r}}{d \zeta^2} + \frac{d \bm{a}}{d \zeta} =  \left( \frac{d \bm{r}}{d \zeta}  \cdot \nabla \right) \bm{a} + \frac{d \bm{r}}{d \zeta} \times (\nabla\times \bm{a}) \mathrm{,}
	\label{eq:lorentz_zeta}
\end{align}
which $\zeta=z-t$ is the proper time. $\bm{a}$ the normalized laser vector potential and $\bm{r}=(x,y,z)$ is the vector of the electron trajectory. The laser pulse propagates in $\hat z$-direction. The momentum is then calculated by $\bm{p} = -d \bm{r}/d\zeta$.
Here, we use dimensionless units (unless the units are explicitly specified) by normalizing time to the laser frequency $\omega_l^{-1}$, length to laser wavenumber $k_l^{-1}$, the electromagnetic fields to $m_e c \omega_l / e$. The plasma density $n_e$ is normalied by the critical density with repect to the laser wavelength $\lambda_l=0.8~\si{\mu m}$.
We consider the laser pulse linearly polarized in $\hat{x}$-direction, $\bm{a}= \hat{x} a_x(x,y,\zeta)$.
The vector potential of the laser field can be expanded at its initial position $(x_0,y_0, z_0)$ as $a_x(X,Y,\zeta) \simeq a_0 f(\zeta) (1 - \mu X + \nu X^2) (1 - \mu Y + \nu Y^2)$ where $X=x-x_0$, $Y=y-y_0$ and $Z=z-z_0$ presenting the offset of the electron from its initial position before the laser pulse arrives. By thinking about the sysmmetry in $\hat x$ and $\hat y$ directions, the coefficients are defined as $\mu= \partial_s a_{x}(x_0,y_0,\zeta)/ a_{x}(x_0,y_0,\zeta) \sim 1/w_0$ and $\nu = \partial_s^2 a_{x}(x_0,y_0,\zeta)/ 2 a_x(x_0,y_0,\zeta)\sim 1/w_0^2$, where the symbol $s$ denotes either $x$ or $y$. Since $\nu \ll \mu \ll 1$, the higher orders are not important.The expansion of $a_x(X,Y,\zeta)$ is reasonable for the paraxial beam, such as Gaussian pulse.  
The transverse momentum is calculated by $p_x=-dX/d\zeta$ or $p_y=-dY/d\zeta$.
With the Coulomp gauge, $\nabla \cdot \bm{a}=0$, we can get $a_z(X, Y, \zeta) = -\int_{\infty}^{\zeta}\partial_X a_x d\zeta' = -a_0 g_1(\zeta) \left( -\mu + 2 \nu X \right) (1 - \mu Y + \nu Y^2)$, where $g_1(\zeta)=\int_{-\infty}^\zeta f(\zeta') d\zeta'$.

\begin{figure}
	\centering
	\includegraphics[width=0.5\textwidth]{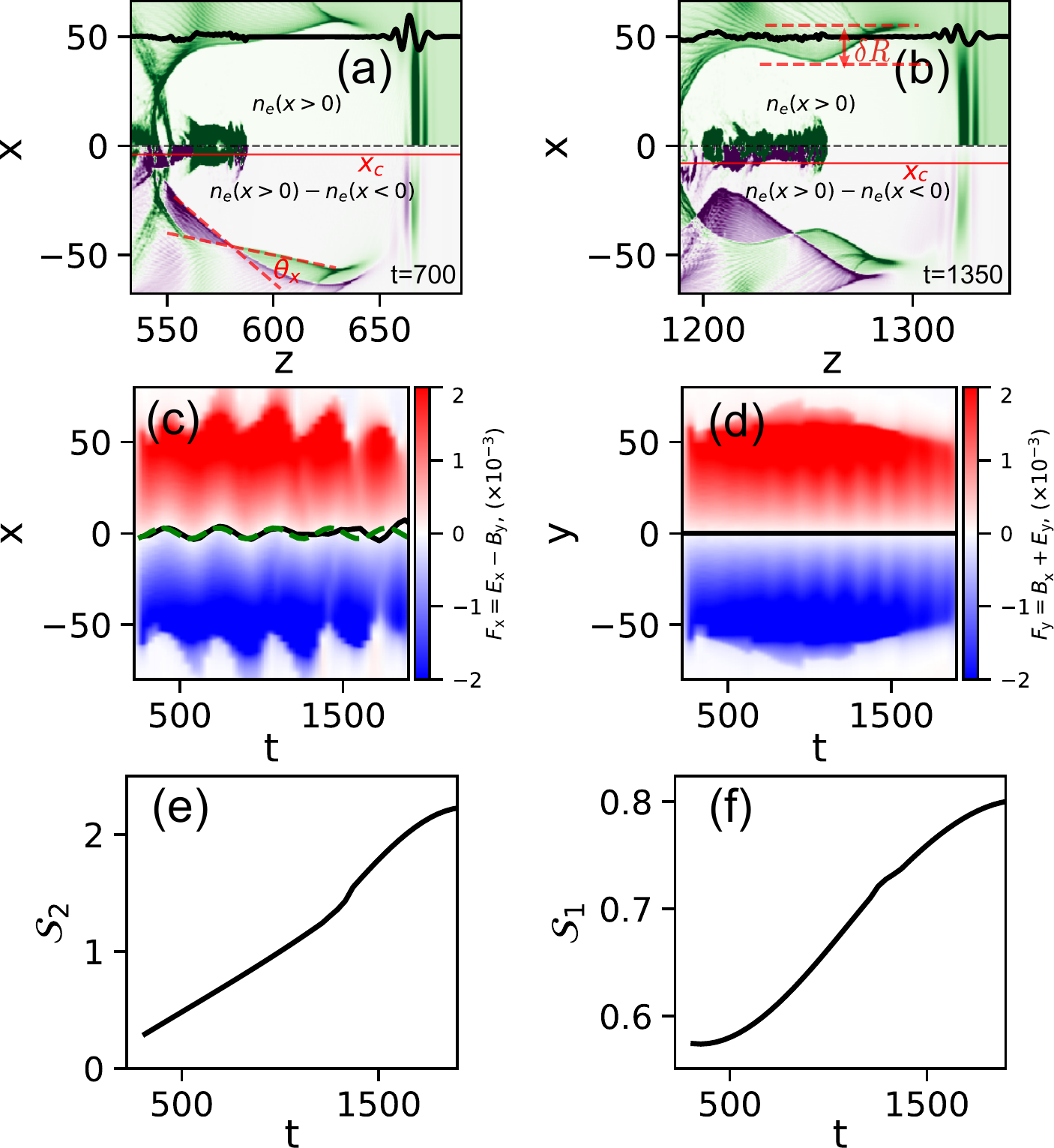}
	\caption{Plasma bubble centroid and sheath oscillations calculated from 3D PIC. (a) and (b) Distribution of plasma electron density and transverse difference at two times, $t=700$ and $t=1350$ respectively. The density difference between the upper ($x>0$) and lower ($x<0$) planes is calculated as $n_{\text{diff}} = n_e(x>0)-n_e(x<0)$. The bubble centroids $x_c$ are indicated by the dashed red lines. The solid black lines present the onaxis($x=0$) electric field $E_x$.
	(c) and (d) Evolution of transverse forces $F_x$ and $F_y$ experienced by the trapped electron inside bubble. The centroid trajectory of the bubble is calculated from PIC (solid black) and theory with Eq.~\eqref{eq:btilt_angle} (dashed green in (c)).
	(e) and (f) Evolution of the defined functions $\mathcal{S}_1$ and $\mathcal{S}_2$ calculated with the onaxis laser field obtained from PIC.}
	\label{fig:BCO_BSO}
\end{figure}

For the wide laser pulse, $w_0 \gg 1$, $X$ and $Y$ can be perturbed with respect to the small parameter $\mu$~\cite{Nerush2009carr} as $X = \sum_{i=0}^{\infty} X^{(i)}$ and $Y = \sum_{i=0}^{\infty} Y^{(i)}$ where $i$ is integer. By inserting this expansion, $a_x(X,Y,\zeta)$ and $a_z(X,Y,\zeta)$ into Eq.~\eqref{eq:lorentz_zeta}, we can get the equations for different orders. 
%See more details in Supplementary~\cite{}.
%%
The $0^{\text{th}}$ order $(X^{(0)}, Y^{(0)},Z^{(0)})$ , which presents the solution of spatial plane wave,  indicates no momentum gain of electrons and therefore does not contribute to the generation of plasma bubble.
The higher order $i\geq 1$ is given as 
\begin{align} 
\label{eq:d2Xi}	
\frac{d^2 X^{(i)}}{d\zeta^2}  &= a_0 f'(\zeta) \bigg[ \mu (X^{(i-1)} + Y^{(i-1)})  \\
& - \sum_{m+n=i-2}(\nu X^{(m)} X^{(n)} + \nu Y^{(m)} Y^{(n)} \nonumber \\
& + \mu^2 X^{(m)}Y^{(n)})\bigg] + 2a_0 g_1(\zeta) \nu \frac{dZ^{i-2}}{d \zeta} + \mathcal{O}_X^{(i)}  \mathrm{,} \nonumber
\end{align} 
and 
\begin{align}
\label{eq:d2Yi}
	\frac{d^2 Y^{(i)}}{d \zeta^2} & = a_0 f(\zeta)\bigg[ \mu \frac{dX^{(i-1)}}{d \zeta} + \sum_{m+n=i-2} \frac{d X^{(m)}}{d\zeta}  \\
	&  \cdot \bigg( -\mu^2 X^{(n)}+ \nu Y^{(n)} \bigg) + \mu^2 \frac{dZ^{(i-2)}}{d \zeta} \bigg] + \mathcal{O}_Y^{(i)} \mathrm{,} \nonumber
\end{align}
where $\mathcal{O}_X^{(i)}$ and $\mathcal{O}_Y^{(i)}$ denote the additional terms of higher orders, e.g. $i>3$ in $\hat x$- and $\hat y$-directions respectively. These terms are small and not important.
As seen from the iterative relation in Eq.~\eqref{eq:d2Xi} and \eqref{eq:d2Yi} that the electron dynamics in $\hat x$-direction dependes on its lower-order offset either in $\hat x$- or $\hat y$-direction. But in $\hat y$-direction it is only on its lower-order momentum in $\hat x$-direction. 
Since generally $X^{(i)} \gg dX^{(i)}/d\zeta$, this elucidates why some effects are only significant in the laser polarization direction~\cite{Ma2020pol, GallardoGonzalez2018aa,Feng2020aa, Seidel2022pointing}.
The $i^{\text{th}}$ order transverse momentum of an electron can be approximated in the sum of different oscillation modes as 
\begin{align}
\label{eq:psi_zeta}
	& p_S^{(i)}(\zeta) \sim a_0 \left(\frac{a_0 \mu}{2} \right)^i  	\cdot \\ 
	& \begin{cases}
		\sum_{j=0}^{\infty} C_{s,j} \cos[2j (\zeta+\phi_{CEP})] & i=2k-1 \\
		\sum_{j=0}^{\infty} C_{s,j} \cos[(2j+1) (\zeta+\phi_{CEP})] & i=2k \\
	\end{cases} \nonumber
\end{align}
where the symbol $S$ denotes either $X$ or $Y$ and $k$ is a positive integer. The longitudinal profile of the laser field is assumed by $f(\zeta)=h(\zeta) \sin(\zeta+\phi_{CEP})$ with the envelope function $h(\zeta)$ and the CEP phase $\phi_{CEP}$. 
The coefficient $C_{S,j} \sim 1/2^{j-1}\leq 1$ and $(a_0 \mu /2) < 1$ for the current laser pulse used in LWFA experiments.
The higher orders is scaled as $P_S^{(i)}\propto a_0^{i+1}/(2w_0)^i$ for $i^{\text{th}}$ order and is negligible. Here, we only consider the $1^{\text{th}}$ and $2^{\text{th}}$ orders. 
%%%%%%%%%%%%%%%%%%

In the 2D plasma system with two electrons of initial positions $x_0$ and $-x_0$ symmetrical with respect to the laser axis $x=0$, it is easy to see that $p_x^{(1)}(-x_0) + p_x^{(1)}(x_0)=0$ and $p_x^{(2)}(-x_0) + p_x^{(2)}(x_0)= 2p_x^{(2)}(x_0)$. 
Noted that these two relations are a simplified version of an initially homogeneous plasma and it should be valid to describe the collective plasma response.
The $1^{\text{st}}$ relation illustrates the plasma response to the laser field with transverse symmetry.  The non-oscillatory component in Eq.~\eqref{eq:psi_zeta} as $m=0$ results in the monotonic increase of the momentum gain during the laser plasma scattering, which determines the transverse dimension of plasma bubble.
The $2^{\text{nd}}$ presents the asymmetric response, which indicates the tilted plasma bubble with its centroid deviating from the laser axis as shown in Fig.~\ref{fig:BCO_BSO}(a) and (b) and as discussed in Ref.~\cite{Nerush2009carr}. 
The phase-dependent oscillation components in Eq.~\eqref{eq:psi_zeta} further indicate that the bubble sheath and the cetroid can oscillate periodically around the laser axis if the phase shifts, which we define here as BSO and BCO effects respectively. The phase variation can be caused by the depersion or pulse etching in the plasma. 
As a result, the transverse force experienced by the trapped electrons is disturbed. 
In the following, the BCO and BSO effects are estimated quantitatively. 
%%

%%%%%%%%%%%%%%%%%
%%
The first order presents the ponderomotive scattering and its momentum is calculated by
\begin{align}
	\label{eq:px1}
	p_x^{(1)} (\zeta) & = p_y^{(1)} (\zeta) \\
	&  = -\frac{a_0^2 \mu}{2} \int_{\infty}^{\zeta}h^2(\zeta^{\prime})[1-\cos(2\zeta' + 2\phi_{CEP})]d\zeta' \mathrm{,} \nonumber 
\end{align} 
which determines the transverse bubble radius $R$ by collectively balancing with the space-charge field.
The oscillatory component in Eq.~\eqref{eq:px1} contributes only when the envelope $h(\zeta)$ becomes temporally asymmetric and short. It therefore cannot be removed by averaging during integration. 
Since the perturbation of the bubble radius, $\delta R$, is due to the variation of the first-order momentum, it can be estimated by the ratio of the oscillatory to non-oscillatory components in Eq.~\eqref{eq:px1} as,
\begin{align}
	\frac{\delta R}{R_0} \simeq \mathcal{S}_1(\zeta_{\tau}) \cos(2 \omega_w t + \phi_{BSO})\mathrm{,}
	\label{eq:dR_R}
\end{align}
where $R_0$ is the bubble radius due to the non-oscillation component and $\zeta_{\tau}$ is the laser pulse duration. $\phi_{BSO}$ is a trivial phase shift during the laser-plasma interaction. 
In the matched bubble, $R_0 \sim \gamma_p \sqrt{a_0}$ where $\gamma_p=k_l/k_p$~\cite{Lu2007gen}. 
The frequency $\omega_w$ can be calculated from the evolution of the pulse front phase $\zeta_f$ as $\omega_w=d \zeta_f /dt \simeq (v_{ph}-v_g+v_{etch}) \simeq 2/\gamma_p^2$ with $v_{ph}$, $v_g$ and $v_{etch}$ are phase, group and etching velocity of the laser pulse respectively.
Here $\mathcal{S}_1(\zeta_{\tau})$ measures the temporal asymmetry of the laser pulse which evolves as the laser pulse being etched and can be defined as
\begin{equation}
	\mathcal{S}_1(\zeta_{\tau}) = \int_{\infty}^{\zeta} h^2(\zeta^{\prime}) \cos(2\zeta')d \zeta' / \int_{\infty}^{\zeta} h^2(\zeta^{\prime}) d \zeta' \mathrm{.}
	\label{eq:S1}
\end{equation} 
It is easy to seen that $\mathcal{S}_1(\zeta_{\tau})$.

As shown in Fig.~\ref{fig:BCO_BSO}(f), $\mathcal{S}_1(\zeta_{\tau})$ increases as the laser pulse propagating in the plasma, which is calculated from a fully 3D PIC simulation. 
The simulations are carried out by SMILEI~\cite{derouillat_smilei_2018}. The grid resolution is $16 \times 16 \times 24$ per laser wavelength in $\hat x$-, $\hat y$- and $\hat z$-direction respectively. 
The laser pulse initially has a temporally asymmetric profile with one cycle of the front and two cycles of the tail and $a_0=6$, $w_0=47$. The plasma is initially set as the natural Helium gas of density $n_e=0.01$ and there are 4 micro-particles in each numerical cell.
%See more details on the simulation method in Supplementary\cite{}.
%%
The bubble with deformed sheath is shown in Fig.~\ref{fig:BCO_BSO} (a) and (b) for different time instants, where $\mathcal{S}_1\simeq 0.55$ and $0.7$ respectively. 
With the evolving phase, BSO with frequency $2\omega_w$ can occurs in both $\hat x$- and $\hat y$-directions as shown in Fig.~\ref{fig:BCO_BSO}(c) and (d). 
As seen in Eq.~\eqref{eq:px1}, BSO occurs both in $\hat x$- and $\hat y$-directions. 
Note that, in $\hat x$-direction, BSO is relatively vague in visualization due to the strong BCO effect as discussed in the following.
%%
%%%%%%%%%%%%%%%%%%%%%

The even orders in Eq.~\eqref{eq:psi_zeta} are oscillating and have the lowest frequency $\omega_l$ as $m=0$. The dominant  order contributing to the plasma bubble is the second with the momentum as, in $\hat x$-direction: $p_x^{(2)} = a_0^3 \mu^2[C_0 \cos(\zeta+\phi_{CEP}) + C_1 \cos(3\zeta+3\phi_{CEP})]$. This results in a tilt angle of plasma bubble as
\begin{align}
	\theta_x \simeq \frac{2 p_x^{(2)}}{p_x^{(1)}} \simeq a_0 \mu\frac{\int f(\zeta^{\prime})^2 g_1(\zeta')d \zeta}{\int f(\zeta')^2d\zeta'}=a_0 \mu \mathcal{S}_2(\zeta)  \mathrm{,}
	\label{eq:btilt_angle}
\end{align}
where $\mathcal{S}_2(\zeta)$ measures sharpness of the laser envelope.
The tilted bubble is shown in Fig.~\ref{fig:BCO_BSO}(a) where the PIC result is $\theta_{x,PIC}\simeq 0.28~\si{rad}$ which is comparable with the theoretical estimation $\theta_{x}\simeq 0.2~\si{rad}$. In theoretical calculation, we use $\mathcal{S}_2=1.0$ and the beam centered at $x_c=1.5$ obtained from PIC.
%%
%The dependence of $\mu$ indicates that the tilt angle replies on the initial position of the electrons and $\theta_x$ in Eq.~\eqref{eq:btilt_angle} calculated along the beam centroid trajectory.
%%
Since $|X^{(1)}| \gg |d X^{(1)}/d\zeta|$, and then $p_x^{(2)} \gg p_y^{(2)}$ or $\theta_x \gg \theta_y$, the bubble tilting effect in the $\hat y$-direction is negligible as seen in Fig.~\ref{fig:BCO_BSO}(d) where the bubble centroid trajectory is well along the laser axis. This agrees with the result in~\cite{Nerush2009carr}.
The tilted bubble evolves over time as BCO effect.
The amplitude can be estimated by how the bubble centroid derivates from the laser axis in the transverse bubble extent, as $a_c \sim \theta_x R_0 \simeq a_0 R_0 \mathcal{S}_2(\zeta)/w_0$.
The centroid trajectory of the bubble in $\hat x$-direction can be written as $x_c(t)\simeq a_c \sin(\omega_w t + \phi_{BCO})$ where $\phi_{BCO}$ is a trivial phase shift during the laser-plasma interaction.
The analytical estimate of $x_c$ agrees well with the PIC result as shown in Fig.~\ref{fig:BCO_BSO}(c) by comparing the dashed green (theory) and solid black (PIC) lines. 
%where the amplitude is quasi-stable, because the pulse depletion counteracts growth of $\mathcal{S}_1$.
%%
Since $\delta R/R_0$ oscillates two times faster than $x_c$, the BCO effect should be easier to become significant than BSO which agrees with most of experiments~\cite{Seidel2022pointing,Ferri2016aa, Corde2013aa}.

%%%%%%%%%%%%%%%%%%
The temporal asymmetric profile can also lead to BSO and BCO effects, such as laser frequency chirp which may have significant effects on plasma bubble dynamics~\cite{Zhang2012eff, Pathak2012eff, Sohbatzadeh2013gro}. For example, considering a linear chirp $\tilde{\omega}_l=1+0.48 \zeta$ and from Eq.~\eqref{eq:S1} and \eqref{eq:btilt_angle} we can calculate $\mathcal{S}_1=0.4$ and $\mathcal{S}_2=1.6$ with a Gaussian laser pulse of the root mean square (RMS) pulse duration $\tau=6$. $\mathcal{S}_1$ and $\mathcal{S}_2$ can be calculated in a similar way for the other types of chirp, e.g. non-linear chirp. This feature allows us to actively control the dynamics of the plasma bubble  and then the trapped electrons, as discussed below.

%%%%%%%%%%%%%%

The transverse force experienced by the trapped electron in this evolving plasma bubble is not linear and can be written with the centroid trajectory and sheath deformation as  $F_x \simeq K (x - x_c)/2 \gamma_p^2$, where the field strength of the bubble is $K=1+\delta R/R_0 = 1 + \mathcal{S}_1(\zeta)\cos(2 \omega_w t + \phi_{BSO}) $. The trapped electron beam undergoes the betatron oscillation inside the plasma bubble and its centroid trajectory can be described by the equation of motion, e.g. in $\hat x$-direction as
\begin{align}
	\frac{d^2 x}{dt^2} + \alpha \frac{dx}{dt}  + K \omega_{\beta}^2 x = K \omega_{\beta}^2x_c(t) \mathrm{,}
	\label{eq:momentum_equation}
\end{align}
where $\omega_{\beta}=1/\gamma_p \sqrt{2\gamma}$ with $\gamma$ Lorentz factor of the electron,
$\alpha=\dot{\gamma}/\gamma$ represents the damping effect due to acceleration. $\omega_{\beta}$ decreases with acceleration. The BCO effect acts as an external driving force and can lead to the linear resonance. The BSO effect modulates the betatron frequency and can lead to the parametric resonance. 
In $\hat y$-direction, since BCO effect is negligible, the equation of motion is same as Eq.~\eqref{eq:momentum_equation} without the an external driving force on right-hand-side(r.h.s).
%%%

\begin{figure}
	\centering
	\includegraphics[width=0.5\textwidth]{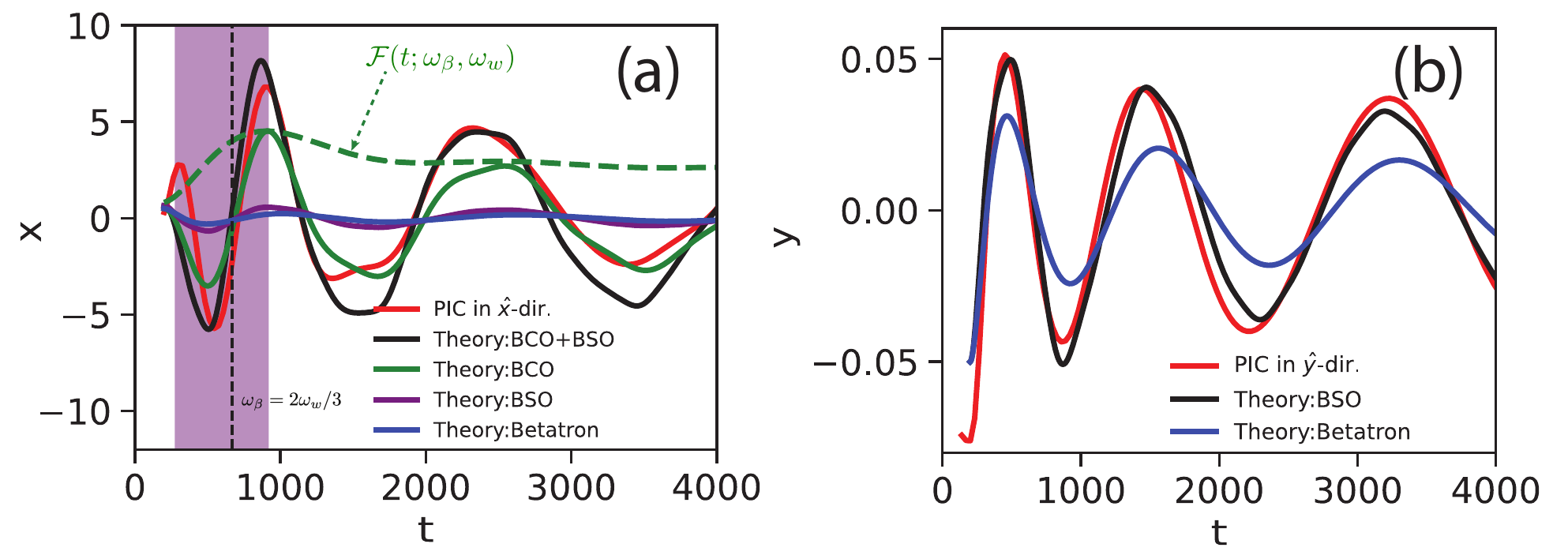}
	\caption{Centroid trajectories of the trapped electron beam in (a): $\hat x$- and (b):$\hat y$-directions, obtained from PIC simulation (red) and theory. The theoretical trajectories are calculated with and without BCO and BSO effects in Eq.~\eqref{eq:momentum_equation}. The vertical dashed line shows the time when $\omega_{\beta}=2\omega_w/3$ is satisfied The colored region in (a) shows the time duration at which the parametric amplification is satisfied according to Eq.~\eqref{eq:KE_condition}. The green dashed line is plotted for the function $\mathcal{F}(t,\omega_{\beta},\omega_w)$ with the laser and plasma parameters used in PIC. The initial plasma density used in these simulations is $n_e=0.001$.}
	\label{fig:trajectory}
\end{figure}

%%%%%%%%%%%%%%%%%%%%
We first consider the case where only BCO is significant, as discussed recently in several papers~\cite{Huijts2021ide, Liu2023bub, Zhang2022car,Kim2023pol}.
The solution of Eq.~\eqref{eq:momentum_equation} is given easily by $x(t) \simeq A(t) R(t) \left[ x_{\beta}(t) + x_D(t) \right ]$, where $x_{\beta}(t)=a_{\beta 0} \cos(\psi_{\beta} + \Phi_{\beta 0})$ presents pure betatron oscillation with amplitude $a_{\beta 0}=\sqrt{x_0^2+2\gamma_0\Theta_0^2}$, depending on the initial injection offset $x_0$ and angle $\Theta_0$ of electron into the bubble, and the initial betatron phase $\Phi_{\beta 0}=\arctan(x_0/2\gamma_0 \Theta_0)$. 
Betatron phase is given as $\psi_{\beta}(t)=\int_{t_0}^t \sqrt{K}\omega_{\beta}(t')dt'$. $A(t)=e^{-\int_{t_0}^t \alpha(t')/4 dt'} =(\gamma_0/\gamma)^{1/4}$ and $R(t)=(K_0/K)^{1/4}$ with $K_0=K|_{t=t_0}$ and $t_0$ is the trapping time. $x_D(t)=2\omega_{\beta 0} K_0 \int_{t_0}^{t} A(t^{\prime}) R(t^{\prime})^{5}  x_c(t^{\prime}) \sin[\psi_{\beta}(t) - \psi_{\beta}(t^{\prime})]dt^{\prime}$ is the BCO-driven solution where $\omega_{\beta 0}$ is the initial betatron frequency. 
The amplitude can be approximated by $|x_D(t)| \simeq \omega_{\beta 0} K_0 a_c \mathcal{F}(t;\omega_{\beta},\omega_w)/2$
with the characteric function defined as
\begin{align}
	\mathcal{F}(t;\omega_{\beta},\omega_w) \simeq \bigg| \int_{t_0}^{t} A(t^{\prime})R(t^{\prime})^{-5} e^{i \int (\omega_{\beta}-\omega_w) dt''}dt^{\prime} \bigg| \mathrm{.}
	\label{eq:Ft}
\end{align} 
$\mathcal{F}(t;\omega_{\beta},\omega_w)$ implies a linear resonance as $\omega_{\beta} \to 2\omega_w/3$ found numerically.
Here, the pure betatron oscillation is not important because the amplitude $a_{\beta 0}$ is much smaller than the amplified $|x_D|$. As shown in Fig.~\ref{fig:trajectory}(a), the amplitude of the pure betatron oscillation (blue) is 1 order smaller than the linear-resonance-amplified betatron oscillation (green), whose amplitude is characterized by $\mathcal{F}(t;\omega_{\beta},\omega_w)$ (dashed green).
In the $\hat y$-direction in Fig.~\ref{fig:trajectory}(b), the electron beam undergoes a very small amplitude betatron oscillation, which is only weakly amplified by the BSO effect, as discussed later.
%%
%See more details of the PIC in Supplementary.
%%%%%%%%%%%%

\begin{figure}
	\centering
	\includegraphics[width=0.4\textwidth]{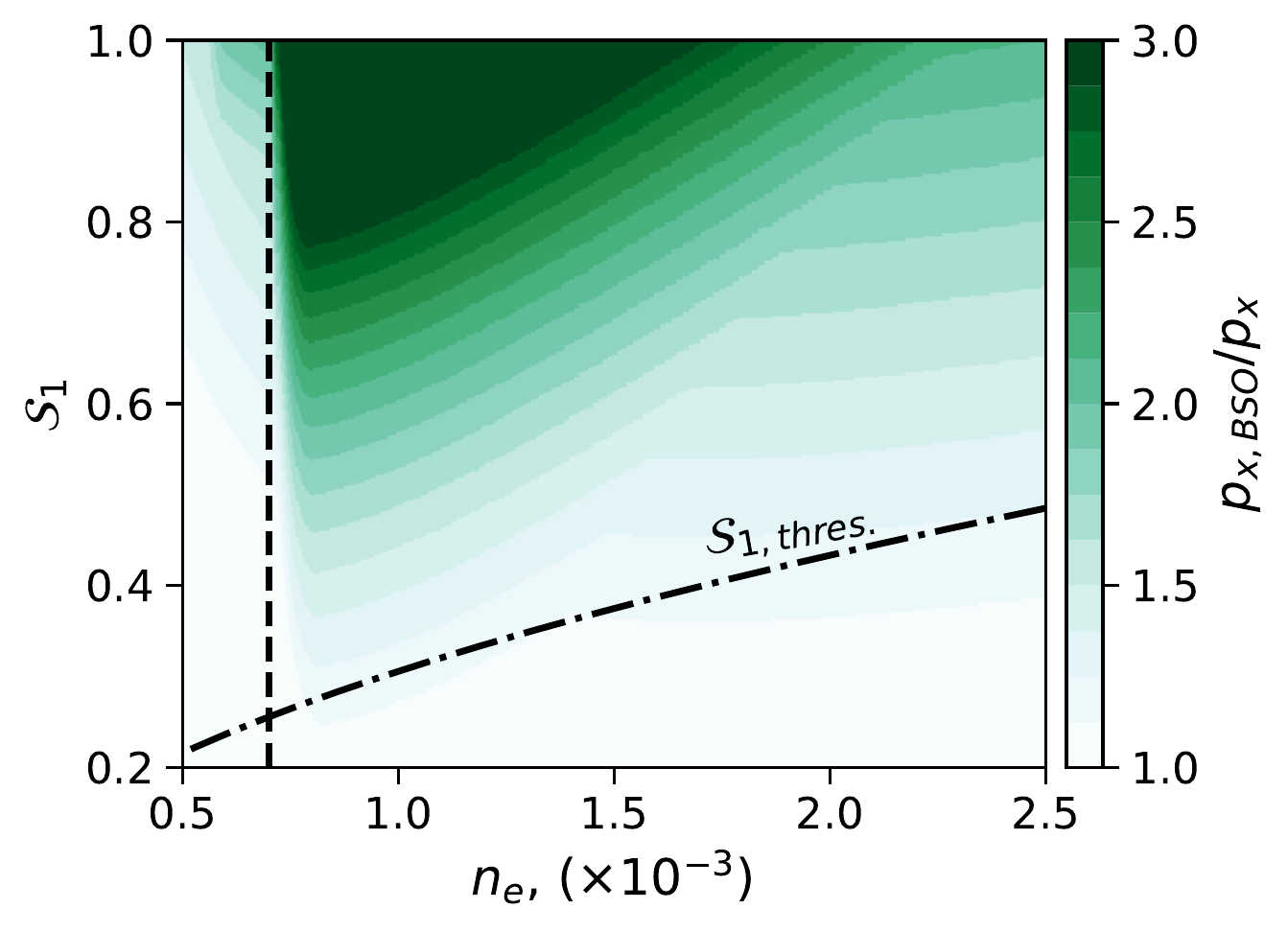}
	\caption{Numerical result: Ratio of the peak of the transverse momentum $p_x$ with and without BSO effect. The dot-dashed line shows the theoretical threshold calculated by the minimum of Eq.~\eqref{eq:KE_condition} for parametric amplification to occur.
	The dashed line shows the position where the plasma density is $n_e=0.7\times 10^{-3}$.
	The laser and electron parameters are $a_0=6, w_0=47, \gamma_0=20$ and $x_a=0.1$.}
	\label{fig:parametric_diag}
\end{figure}

Next, when BSO becomes strong to modulate the betatron frequency in Eq.~\eqref{eq:momentum_equation}, the centroid dynamics of the electron beam can show a parametric amplification with the betatron frequency $\omega_{\beta}$ close to $\omega_w$~\cite{cartmell_introduction_1990}.
Assuming that $\mathcal{S}_1(\zeta)$ shifts slowly in comapring with betatron oscillation as seen in Fig.~\ref{fig:BCO_BSO}(f), the solution can be approximated as $x(t) \simeq x_a e^{\Gamma(t)} \sin(\omega_w t)$ where $x_a$ is the initial amplitude and $\Gamma(t) = \int_0^t [\eta(t') - \alpha(t')]dt'$ with $\eta(t) = (\omega_{\beta}^2/2\omega_w) \sqrt{4 \mathcal{S}_1^2 -\left( \omega_w^2/\omega_{\beta}^2 - 1 \right)^2}$. 
The solution presents the exponential growth if $\eta(t) - \alpha(t)>0$. 
This indicates a parametric amplification of the betatron oscillation when the laser pulse becomes highly asymmetric to exceed the threshold as   
\begin{equation}
\centering
\mathcal{S}_1 > \frac{1}{2\omega_{\beta}^2}\sqrt{4 \alpha^2 \omega_w^2 + \left(\omega_w^2 - \omega_{\beta}^2 \right)^2}  \mathrm{,}
	\label{eq:KE_condition}
\end{equation}
or  it is equal to set a frequency band as $|\Delta \omega| = |\omega_{\beta} - \omega_w| < |1-\rho| \omega_w$
where $\rho=\sqrt{(M-1)/(\mathcal{S}_1^2/4 - 1)}/2$ and $M=\sqrt{1 + (\mathcal{S}_1^2/4 - 1)(1 + 4 \alpha^2/\omega_w^2)}$. 
%%
%Practically, $0.5 \leq |1-\rho|\leq 1$ with  the laser and plasma parameters used in current LWFA.
%%
The parametric oscillation depends on the initial amplitude $x_a$, which means that the BSO itself can not drive the oscillation unless it couples with the other process, such as the betatron oscillation. With BCO in $\hat x$-direction the betatron oscillation can be strongly enhanced as shown in Fig.~\ref{fig:parametric_diag}(a) where the analytical result agrees well with PIC by comparing the black and red lines.
The parametric amplification also occurs in the $\hat y$-directions as shown in Fig.~\ref{fig:trajectory}(b), where the enhancement of the betatron amplitude can be seen by comparing the red and blue lines. 
However, the betatron amplitude in $\hat y$-direction is much smaller, e.g. 2 orders lower, due to the absence of the linear amplification driven by BCO.
Of course, the parametric amplification can also be significant if the amplitude of the pure betatron oscillation $a_{\beta 0}$ could be large due to the large initial amplitude induced, for example, by CEP-controlled injection~\cite{Kim2021sub, Huijts2022wave}.
The parametric amplification is limited by the threshold relation in Eq.~\eqref{eq:KE_condition} which indicates that this process can only persist for a limited time. As a result, the enhancement of the betatron oscillation is generally moderate unless the special strategies are considered to bring the laser plasma parameters into the strong amplification region as shown in Fig.~\ref{fig:parametric_diag}.

%%%%%
The parameter regime for the amplification to happen is shown in Fig.~\ref{fig:parametric_diag}. 
The diagram is calculated from Eq.~\eqref{eq:momentum_equation} for an electron beam propagating over $t=4000$ by the ratio of maximum of the transverse momentum $p_{x,BSO}$ with BSO and BCO considered to that $p_{x}$ only with BCO. 
It can be seen that, with BSO effect, the betatron oscillation is amplified, especially in a certain range of $n_e$ and $\mathcal{S}_1$.
It is interesting to note that there is a density threshold, e.g. $n_e\simeq 0.7\times 10^{-3}$ with the parameters used, below which the parametric amplification becomes weak. 
For higher densities, e.g. $n_e>2.0\times 10^{-3}$, the larger value of $\mathcal{S}_1$ is needed. This can be explained by the parametric resonance condition $\omega_{\beta}=\omega_w$ which gives the early resonance time and narrow frequency band for high plasma density. As a result, the parametric amplification is inefficient due to small $x_a$. 
%%
%These features provide us with a method to manipulate the parametric resonance and then the betatron dynamics.

%%

%In conclusion, the hign-intensity laser driven plasma bubble shows the phase-dependent hosing which is instrinsic in transverse directions. However, it becomes significant only in the polarization direction of the laser pulse due to the strong coupling between BCO and BSO. As a result, it substantially amplfy the betatron oscillation of the trapped electron beam in this direction.
%%
The phase-dependent hosing instability of the plasma bubble can lead to the resonant amplification of the betatron oscillation and differs from the commonly accepted mechanism driven by the coupling between the laser and the transverse focusing field~\cite{Cipiccia2011ga, Nemeth2008laser, Curcio2015re, Horny2020att}.
It is therefore responsible for the primary qualities of the beam accelerated from LWFA, such as the shot-to-shot beam pointing instability, beam phase space or emittance.
This effect depends on the asymmetry of the temporal laser profile, such as few-cycle pulse, multi-cycle pulse with a sharp front, frequency chirped pulse, or other deliberately constructed asymmetric pulse.  
For example, a triangle pulse with highly asymmetrical envelope can be prepared by passing a Gaussian pulse through a $\si{nm}$ foil. The more complex laser pulse profile should also be possible by combining several pulses.
This provides a novel mechanism to actively manipulate the betatron oscillation for applications, such as to optimize the electron acceleration or betatron radiation.

\begin{acknowledgements}
This work is supported by the National Key R\&D Program of China (Grant No. 2022YFA1603200, 2022YFA1603201); the National Natural Science Foundation of China (Grant Nos. 12135001, 11825502); the Strategic Priority Research Program of the Chinese Academy of Sciences (Grant No. XDA25050900). 
Bin Liu acknowledges the support of Guangdong High Level Innovation Research Institute Project, Grant No. 2021B0909050006. 
B.Q. acknowledges support from the National Natural Science Funds for Distinguished Young Scholars (Grant No.11825502). 
The authors gratefully acknowledge that the computing time is provided by the Tianhe-2 supercomputer at the National Supercomputer Center in Guangzhou and the Gauss Centre for Supercomputing e.V. (www.gauss-centre.eu) through the John von Neumann Institute for Computing (NIC) on the GCS Supercomputer JUWELS at Jülich Supercomputing Centre (JSC).
\end{acknowledgements}

\bibliography{pbaref.bib}

\begin{thebibliography}{46}%
\makeatletter
\providecommand \@ifxundefined [1]{%
 \@ifx{#1\undefined}
}%
\providecommand \@ifnum [1]{%
 \ifnum #1\expandafter \@firstoftwo
 \else \expandafter \@secondoftwo
 \fi
}%
\providecommand \@ifx [1]{%
 \ifx #1\expandafter \@firstoftwo
 \else \expandafter \@secondoftwo
 \fi
}%
\providecommand \natexlab [1]{#1}%
\providecommand \enquote  [1]{``#1''}%
\providecommand \bibnamefont  [1]{#1}%
\providecommand \bibfnamefont [1]{#1}%
\providecommand \citenamefont [1]{#1}%
\providecommand \href@noop [0]{\@secondoftwo}%
\providecommand \href [0]{\begingroup \@sanitize@url \@href}%
\providecommand \@href[1]{\@@startlink{#1}\@@href}%
\providecommand \@@href[1]{\endgroup#1\@@endlink}%
\providecommand \@sanitize@url [0]{\catcode `\\12\catcode `\$12\catcode `\&12\catcode `\#12\catcode `\^12\catcode `\_12\catcode `\%12\relax}%
\providecommand \@@startlink[1]{}%
\providecommand \@@endlink[0]{}%
\providecommand \url  [0]{\begingroup\@sanitize@url \@url }%
\providecommand \@url [1]{\endgroup\@href {#1}{\urlprefix }}%
\providecommand \urlprefix  [0]{URL }%
\providecommand \Eprint [0]{\href }%
\providecommand \doibase [0]{https://doi.org/}%
\providecommand \selectlanguage [0]{\@gobble}%
\providecommand \bibinfo  [0]{\@secondoftwo}%
\providecommand \bibfield  [0]{\@secondoftwo}%
\providecommand \translation [1]{[#1]}%
\providecommand \BibitemOpen [0]{}%
\providecommand \bibitemStop [0]{}%
\providecommand \bibitemNoStop [0]{.\EOS\space}%
\providecommand \EOS [0]{\spacefactor3000\relax}%
\providecommand \BibitemShut  [1]{\csname bibitem#1\endcsname}%
\let\auto@bib@innerbib\@empty
%</preamble>
\bibitem [{\citenamefont {Poder}\ \emph {et~al.}(2018)\citenamefont {Poder}, \citenamefont {Tamburini}, \citenamefont {Sarri}, \citenamefont {Di~Piazza}, \citenamefont {Kuschel}, \citenamefont {Baird}, \citenamefont {Behm}, \citenamefont {Bohlen}, \citenamefont {Cole}, \citenamefont {Corvan}, \citenamefont {Duff}, \citenamefont {Gerstmayr}, \citenamefont {Keitel}, \citenamefont {Krushelnick}, \citenamefont {Mangles}, \citenamefont {McKenna}, \citenamefont {Murphy}, \citenamefont {Najmudin}, \citenamefont {Ridgers}, \citenamefont {Samarin}, \citenamefont {Symes}, \citenamefont {Thomas}, \citenamefont {Warwick},\ and\ \citenamefont {Zepf}}]{Poder2018EXP}%
  \BibitemOpen
  \bibfield  {author} {\bibinfo {author} {\bibfnamefont {K.}~\bibnamefont {Poder}}, \bibinfo {author} {\bibfnamefont {M.}~\bibnamefont {Tamburini}}, \bibinfo {author} {\bibfnamefont {G.}~\bibnamefont {Sarri}}, \bibinfo {author} {\bibfnamefont {A.}~\bibnamefont {Di~Piazza}}, \bibinfo {author} {\bibfnamefont {S.}~\bibnamefont {Kuschel}}, \bibinfo {author} {\bibfnamefont {C.~D.}\ \bibnamefont {Baird}}, \bibinfo {author} {\bibfnamefont {K.}~\bibnamefont {Behm}}, \bibinfo {author} {\bibfnamefont {S.}~\bibnamefont {Bohlen}}, \bibinfo {author} {\bibfnamefont {J.~M.}\ \bibnamefont {Cole}}, \bibinfo {author} {\bibfnamefont {D.~J.}\ \bibnamefont {Corvan}}, \bibinfo {author} {\bibfnamefont {M.}~\bibnamefont {Duff}}, \bibinfo {author} {\bibfnamefont {E.}~\bibnamefont {Gerstmayr}}, \bibinfo {author} {\bibfnamefont {C.~H.}\ \bibnamefont {Keitel}}, \bibinfo {author} {\bibfnamefont {K.}~\bibnamefont {Krushelnick}}, \bibinfo {author} {\bibfnamefont {S.~P.~D.}\ \bibnamefont {Mangles}}, \bibinfo {author} {\bibfnamefont {P.}~\bibnamefont {McKenna}}, \bibinfo {author} {\bibfnamefont {C.~D.}\ \bibnamefont {Murphy}}, \bibinfo {author} {\bibfnamefont {Z.}~\bibnamefont {Najmudin}}, \bibinfo {author} {\bibfnamefont {C.~P.}\ \bibnamefont {Ridgers}}, \bibinfo {author} {\bibfnamefont {G.~M.}\ \bibnamefont {Samarin}}, \bibinfo {author} {\bibfnamefont {D.~R.}\ \bibnamefont {Symes}}, \bibinfo {author} {\bibfnamefont {A.~G.~R.}\ \bibnamefont {Thomas}}, \bibinfo {author} {\bibfnamefont {J.}~\bibnamefont {Warwick}},\ and\ \bibinfo {author} {\bibfnamefont {M.}~\bibnamefont {Zepf}},\ }\bibfield  {title} {\bibinfo {title} {Experimental signatures of the quantum nature of radiation reaction in the field of an ultraintense laser},\ }\href {https://doi.org/10.1103/PhysRevX.8.031004} {\bibfield  {journal} {\bibinfo  {journal} {Phys. Rev. X}\ }\textbf {\bibinfo {volume} {8}},\ \bibinfo {pages} {031004} (\bibinfo {year} {2018})}\BibitemShut {NoStop}%
\bibitem [{\citenamefont {Cole}\ \emph {et~al.}(2018)\citenamefont {Cole}, \citenamefont {Behm}, \citenamefont {Gerstmayr}, \citenamefont {Blackburn}, \citenamefont {Wood}, \citenamefont {Baird}, \citenamefont {Duff}, \citenamefont {Harvey}, \citenamefont {Ilderton}, \citenamefont {Joglekar}, \citenamefont {Krushelnick}, \citenamefont {Kuschel}, \citenamefont {Marklund}, \citenamefont {McKenna}, \citenamefont {Murphy}, \citenamefont {Poder}, \citenamefont {Ridgers}, \citenamefont {Samarin}, \citenamefont {Sarri}, \citenamefont {Symes}, \citenamefont {Thomas}, \citenamefont {Warwick}, \citenamefont {Zepf}, \citenamefont {Najmudin},\ and\ \citenamefont {Mangles}}]{Cole2018EXP}%
  \BibitemOpen
  \bibfield  {author} {\bibinfo {author} {\bibfnamefont {J.~M.}\ \bibnamefont {Cole}}, \bibinfo {author} {\bibfnamefont {K.~T.}\ \bibnamefont {Behm}}, \bibinfo {author} {\bibfnamefont {E.}~\bibnamefont {Gerstmayr}}, \bibinfo {author} {\bibfnamefont {T.~G.}\ \bibnamefont {Blackburn}}, \bibinfo {author} {\bibfnamefont {J.~C.}\ \bibnamefont {Wood}}, \bibinfo {author} {\bibfnamefont {C.~D.}\ \bibnamefont {Baird}}, \bibinfo {author} {\bibfnamefont {M.~J.}\ \bibnamefont {Duff}}, \bibinfo {author} {\bibfnamefont {C.}~\bibnamefont {Harvey}}, \bibinfo {author} {\bibfnamefont {A.}~\bibnamefont {Ilderton}}, \bibinfo {author} {\bibfnamefont {A.~S.}\ \bibnamefont {Joglekar}}, \bibinfo {author} {\bibfnamefont {K.}~\bibnamefont {Krushelnick}}, \bibinfo {author} {\bibfnamefont {S.}~\bibnamefont {Kuschel}}, \bibinfo {author} {\bibfnamefont {M.}~\bibnamefont {Marklund}}, \bibinfo {author} {\bibfnamefont {P.}~\bibnamefont {McKenna}}, \bibinfo {author} {\bibfnamefont {C.~D.}\ \bibnamefont {Murphy}}, \bibinfo {author} {\bibfnamefont {K.}~\bibnamefont {Poder}}, \bibinfo {author} {\bibfnamefont {C.~P.}\ \bibnamefont {Ridgers}}, \bibinfo {author} {\bibfnamefont {G.~M.}\ \bibnamefont {Samarin}}, \bibinfo {author} {\bibfnamefont {G.}~\bibnamefont {Sarri}}, \bibinfo {author} {\bibfnamefont {D.~R.}\ \bibnamefont {Symes}}, \bibinfo {author} {\bibfnamefont {A.~G.~R.}\ \bibnamefont {Thomas}}, \bibinfo {author} {\bibfnamefont {J.}~\bibnamefont {Warwick}}, \bibinfo {author} {\bibfnamefont {M.}~\bibnamefont {Zepf}}, \bibinfo {author} {\bibfnamefont {Z.}~\bibnamefont {Najmudin}},\ and\ \bibinfo {author} {\bibfnamefont {S.~P.~D.}\ \bibnamefont {Mangles}},\ }\bibfield  {title} {\bibinfo {title} {Experimental evidence of radiation reaction in the collision of a high-intensity laser pulse with a laser-wakefield accelerated electron beam},\ }\href {https://doi.org/10.1103/PhysRevX.8.011020} {\bibfield  {journal} {\bibinfo  {journal} {Phys. Rev. X}\ }\textbf {\bibinfo {volume} {8}},\ \bibinfo {pages} {011020} (\bibinfo {year} {2018})}\BibitemShut
  {NoStop}%
\bibitem [{\citenamefont {Turner}\ \emph {et~al.}(2022)\citenamefont {Turner}, \citenamefont {Bulanov}, \citenamefont {Benedetti}, \citenamefont {Gonsalves}, \citenamefont {Leemans}, \citenamefont {Nakamura}, \citenamefont {van Tilborg}, \citenamefont {Schroeder}, \citenamefont {Geddes},\ and\ \citenamefont {Esarey}}]{turner2022strong}%
  \BibitemOpen
  \bibfield  {author} {\bibinfo {author} {\bibfnamefont {M.}~\bibnamefont {Turner}}, \bibinfo {author} {\bibfnamefont {S.}~\bibnamefont {Bulanov}}, \bibinfo {author} {\bibfnamefont {C.}~\bibnamefont {Benedetti}}, \bibinfo {author} {\bibfnamefont {A.}~\bibnamefont {Gonsalves}}, \bibinfo {author} {\bibfnamefont {W.}~\bibnamefont {Leemans}}, \bibinfo {author} {\bibfnamefont {K.}~\bibnamefont {Nakamura}}, \bibinfo {author} {\bibfnamefont {J.}~\bibnamefont {van Tilborg}}, \bibinfo {author} {\bibfnamefont {C.}~\bibnamefont {Schroeder}}, \bibinfo {author} {\bibfnamefont {C.}~\bibnamefont {Geddes}},\ and\ \bibinfo {author} {\bibfnamefont {E.}~\bibnamefont {Esarey}},\ }\bibfield  {title} {\bibinfo {title} {Strong-field {QED} experiments using the {BELLA} pw laser dual beamlines},\ }\href {https://doi.org/10.1140/ep jd/s10053-022-00535-y} {\bibfield  {journal} {\bibinfo  {journal} {The European Physical Journal D}\ }\textbf {\bibinfo {volume} {76}},\ \bibinfo {pages} {1} (\bibinfo {year} {2022})}\BibitemShut {NoStop}%
\bibitem [{\citenamefont {Wang}\ \emph {et~al.}(2021)\citenamefont {Wang}, \citenamefont {Feng}, \citenamefont {Ke}, \citenamefont {Yu}, \citenamefont {Xu}, \citenamefont {Qi}, \citenamefont {Chen}, \citenamefont {Qin}, \citenamefont {Zhang}, \citenamefont {Fang} \emph {et~al.}}]{Wang2021FREE}%
  \BibitemOpen
  \bibfield  {author} {\bibinfo {author} {\bibfnamefont {W.}~\bibnamefont {Wang}}, \bibinfo {author} {\bibfnamefont {K.}~\bibnamefont {Feng}}, \bibinfo {author} {\bibfnamefont {L.}~\bibnamefont {Ke}}, \bibinfo {author} {\bibfnamefont {C.}~\bibnamefont {Yu}}, \bibinfo {author} {\bibfnamefont {Y.}~\bibnamefont {Xu}}, \bibinfo {author} {\bibfnamefont {R.}~\bibnamefont {Qi}}, \bibinfo {author} {\bibfnamefont {Y.}~\bibnamefont {Chen}}, \bibinfo {author} {\bibfnamefont {Z.}~\bibnamefont {Qin}}, \bibinfo {author} {\bibfnamefont {Z.}~\bibnamefont {Zhang}}, \bibinfo {author} {\bibfnamefont {M.}~\bibnamefont {Fang}}, \emph {et~al.},\ }\bibfield  {title} {\bibinfo {title} {Free-electron lasing at 27 nanometres based on a laser wakefield accelerator},\ }\href {https://doi.org/10.1038/s41586-021-03678-x} {\bibfield  {journal} {\bibinfo  {journal} {Nature}\ }\textbf {\bibinfo {volume} {595}},\ \bibinfo {pages} {516} (\bibinfo {year} {2021})}\BibitemShut {NoStop}%
\bibitem [{\citenamefont {Labat}\ \emph {et~al.}(2022)\citenamefont {Labat}, \citenamefont {Cabada{\u{g}}}, \citenamefont {Ghaith}, \citenamefont {Irman}, \citenamefont {Berlioux}, \citenamefont {Berteaud}, \citenamefont {Blache}, \citenamefont {Bock}, \citenamefont {Bouvet}, \citenamefont {Briquez} \emph {et~al.}}]{labat2022seeded}%
  \BibitemOpen
  \bibfield  {author} {\bibinfo {author} {\bibfnamefont {M.}~\bibnamefont {Labat}}, \bibinfo {author} {\bibfnamefont {J.~C.}\ \bibnamefont {Cabada{\u{g}}}}, \bibinfo {author} {\bibfnamefont {A.}~\bibnamefont {Ghaith}}, \bibinfo {author} {\bibfnamefont {A.}~\bibnamefont {Irman}}, \bibinfo {author} {\bibfnamefont {A.}~\bibnamefont {Berlioux}}, \bibinfo {author} {\bibfnamefont {P.}~\bibnamefont {Berteaud}}, \bibinfo {author} {\bibfnamefont {F.}~\bibnamefont {Blache}}, \bibinfo {author} {\bibfnamefont {S.}~\bibnamefont {Bock}}, \bibinfo {author} {\bibfnamefont {F.}~\bibnamefont {Bouvet}}, \bibinfo {author} {\bibfnamefont {F.}~\bibnamefont {Briquez}}, \emph {et~al.},\ }\bibfield  {title} {\bibinfo {title} {Seeded free-electron laser driven by a compact laser plasma accelerator},\ }\bibfield  {journal} {\bibinfo  {journal} {Nature Photonics}\ }\href {https://doi.org/10.1038/s41566-022-01104-w} {10.1038/s41566-022-01104-w} (\bibinfo {year} {2022})\BibitemShut {NoStop}%
\bibitem [{\citenamefont {Pompili}\ \emph {et~al.}(2022)\citenamefont {Pompili}, \citenamefont {Alesini}, \citenamefont {Anania}, \citenamefont {Arjmand}, \citenamefont {Behtouei}, \citenamefont {Bellaveglia}, \citenamefont {Biagioni}, \citenamefont {Buonomo}, \citenamefont {Cardelli}, \citenamefont {Carpanese} \emph {et~al.}}]{pompili2022free}%
  \BibitemOpen
  \bibfield  {author} {\bibinfo {author} {\bibfnamefont {R.}~\bibnamefont {Pompili}}, \bibinfo {author} {\bibfnamefont {D.}~\bibnamefont {Alesini}}, \bibinfo {author} {\bibfnamefont {M.}~\bibnamefont {Anania}}, \bibinfo {author} {\bibfnamefont {S.}~\bibnamefont {Arjmand}}, \bibinfo {author} {\bibfnamefont {M.}~\bibnamefont {Behtouei}}, \bibinfo {author} {\bibfnamefont {M.}~\bibnamefont {Bellaveglia}}, \bibinfo {author} {\bibfnamefont {A.}~\bibnamefont {Biagioni}}, \bibinfo {author} {\bibfnamefont {B.}~\bibnamefont {Buonomo}}, \bibinfo {author} {\bibfnamefont {F.}~\bibnamefont {Cardelli}}, \bibinfo {author} {\bibfnamefont {M.}~\bibnamefont {Carpanese}}, \emph {et~al.},\ }\bibfield  {title} {\bibinfo {title} {Free-electron lasing with compact beam-driven plasma wakefield accelerator},\ }\href {https://doi.org/10.1038/s41586-022-04589-1} {\bibfield  {journal} {\bibinfo  {journal} {Nature}\ }\textbf {\bibinfo {volume} {605}},\ \bibinfo {pages} {659} (\bibinfo {year} {2022})}\BibitemShut {NoStop}%
\bibitem [{\citenamefont {Galletti}\ \emph {et~al.}(2022)\citenamefont {Galletti}, \citenamefont {Alesini}, \citenamefont {Anania}, \citenamefont {Arjmand}, \citenamefont {Behtouei}, \citenamefont {Bellaveglia}, \citenamefont {Biagioni}, \citenamefont {Buonomo}, \citenamefont {Cardelli}, \citenamefont {Carpanese}, \citenamefont {Chiadroni}, \citenamefont {Cianchi}, \citenamefont {Costa}, \citenamefont {Del~Dotto}, \citenamefont {Del~Giorno}, \citenamefont {Dipace}, \citenamefont {Doria}, \citenamefont {Filippi}, \citenamefont {Franzini}, \citenamefont {Giannessi}, \citenamefont {Giribono}, \citenamefont {Iovine}, \citenamefont {Lollo}, \citenamefont {Mostacci}, \citenamefont {Nguyen}, \citenamefont {Opromolla}, \citenamefont {Pellegrino}, \citenamefont {Petralia}, \citenamefont {Petrillo}, \citenamefont {Piersanti}, \citenamefont {Di~Pirro}, \citenamefont {Pompili}, \citenamefont {Romeo}, \citenamefont {Rossi}, \citenamefont {Selce}, \citenamefont {Shpakov}, \citenamefont {Stella}, \citenamefont {Vaccarezza}, \citenamefont {Villa}, \citenamefont {Zigler},\ and\ \citenamefont {Ferrario}}]{Galletti2022stable}%
  \BibitemOpen
  \bibfield  {author} {\bibinfo {author} {\bibfnamefont {M.}~\bibnamefont {Galletti}}, \bibinfo {author} {\bibfnamefont {D.}~\bibnamefont {Alesini}}, \bibinfo {author} {\bibfnamefont {M.~P.}\ \bibnamefont {Anania}}, \bibinfo {author} {\bibfnamefont {S.}~\bibnamefont {Arjmand}}, \bibinfo {author} {\bibfnamefont {M.}~\bibnamefont {Behtouei}}, \bibinfo {author} {\bibfnamefont {M.}~\bibnamefont {Bellaveglia}}, \bibinfo {author} {\bibfnamefont {A.}~\bibnamefont {Biagioni}}, \bibinfo {author} {\bibfnamefont {B.}~\bibnamefont {Buonomo}}, \bibinfo {author} {\bibfnamefont {F.}~\bibnamefont {Cardelli}}, \bibinfo {author} {\bibfnamefont {M.}~\bibnamefont {Carpanese}}, \bibinfo {author} {\bibfnamefont {E.}~\bibnamefont {Chiadroni}}, \bibinfo {author} {\bibfnamefont {A.}~\bibnamefont {Cianchi}}, \bibinfo {author} {\bibfnamefont {G.}~\bibnamefont {Costa}}, \bibinfo {author} {\bibfnamefont {A.}~\bibnamefont {Del~Dotto}}, \bibinfo {author} {\bibfnamefont {M.}~\bibnamefont {Del~Giorno}}, \bibinfo {author} {\bibfnamefont {F.}~\bibnamefont {Dipace}}, \bibinfo {author} {\bibfnamefont {A.}~\bibnamefont {Doria}}, \bibinfo {author} {\bibfnamefont {F.}~\bibnamefont {Filippi}}, \bibinfo {author} {\bibfnamefont {G.}~\bibnamefont {Franzini}}, \bibinfo {author} {\bibfnamefont {L.}~\bibnamefont {Giannessi}}, \bibinfo {author} {\bibfnamefont {A.}~\bibnamefont {Giribono}}, \bibinfo {author} {\bibfnamefont {P.}~\bibnamefont {Iovine}}, \bibinfo {author} {\bibfnamefont {V.}~\bibnamefont {Lollo}}, \bibinfo {author} {\bibfnamefont {A.}~\bibnamefont {Mostacci}}, \bibinfo {author} {\bibfnamefont {F.}~\bibnamefont {Nguyen}}, \bibinfo {author} {\bibfnamefont {M.}~\bibnamefont {Opromolla}}, \bibinfo {author} {\bibfnamefont {L.}~\bibnamefont {Pellegrino}}, \bibinfo {author} {\bibfnamefont {A.}~\bibnamefont {Petralia}}, \bibinfo {author} {\bibfnamefont {V.}~\bibnamefont {Petrillo}}, \bibinfo {author} {\bibfnamefont {L.}~\bibnamefont {Piersanti}}, \bibinfo {author} {\bibfnamefont {G.}~\bibnamefont {Di~Pirro}}, \bibinfo {author} {\bibfnamefont
  {R.}~\bibnamefont {Pompili}}, \bibinfo {author} {\bibfnamefont {S.}~\bibnamefont {Romeo}}, \bibinfo {author} {\bibfnamefont {A.~R.}\ \bibnamefont {Rossi}}, \bibinfo {author} {\bibfnamefont {A.}~\bibnamefont {Selce}}, \bibinfo {author} {\bibfnamefont {V.}~\bibnamefont {Shpakov}}, \bibinfo {author} {\bibfnamefont {A.}~\bibnamefont {Stella}}, \bibinfo {author} {\bibfnamefont {C.}~\bibnamefont {Vaccarezza}}, \bibinfo {author} {\bibfnamefont {F.}~\bibnamefont {Villa}}, \bibinfo {author} {\bibfnamefont {A.}~\bibnamefont {Zigler}},\ and\ \bibinfo {author} {\bibfnamefont {M.}~\bibnamefont {Ferrario}},\ }\bibfield  {title} {\bibinfo {title} {Stable operation of a free-electron laser driven by a plasma accelerator},\ }\href {https://doi.org/10.1103/PhysRevLett.129.234801} {\bibfield  {journal} {\bibinfo  {journal} {Phys. Rev. Lett.}\ }\textbf {\bibinfo {volume} {129}},\ \bibinfo {pages} {234801} (\bibinfo {year} {2022})}\BibitemShut {NoStop}%
\bibitem [{\citenamefont {Schroeder}\ \emph {et~al.}(2010)\citenamefont {Schroeder}, \citenamefont {Esarey}, \citenamefont {Geddes}, \citenamefont {Benedetti},\ and\ \citenamefont {Leemans}}]{Schroeder2010}%
  \BibitemOpen
  \bibfield  {author} {\bibinfo {author} {\bibfnamefont {C.~B.}\ \bibnamefont {Schroeder}}, \bibinfo {author} {\bibfnamefont {E.}~\bibnamefont {Esarey}}, \bibinfo {author} {\bibfnamefont {C.~G.~R.}\ \bibnamefont {Geddes}}, \bibinfo {author} {\bibfnamefont {C.}~\bibnamefont {Benedetti}},\ and\ \bibinfo {author} {\bibfnamefont {W.~P.}\ \bibnamefont {Leemans}},\ }\bibfield  {title} {\bibinfo {title} {Physics considerations for laser-plasma linear colliders},\ }\href {https://doi.org/10.1103/PhysRevSTAB.13.101301} {\bibfield  {journal} {\bibinfo  {journal} {Phys. Rev. ST Accel. Beams}\ }\textbf {\bibinfo {volume} {13}},\ \bibinfo {pages} {101301} (\bibinfo {year} {2010})}\BibitemShut {NoStop}%
\bibitem [{\citenamefont {Schroeder}\ \emph {et~al.}(2016)\citenamefont {Schroeder}, \citenamefont {Benedetti}, \citenamefont {Esarey},\ and\ \citenamefont {Leemans}}]{SCHROEDER2016113}%
  \BibitemOpen
  \bibfield  {author} {\bibinfo {author} {\bibfnamefont {C.}~\bibnamefont {Schroeder}}, \bibinfo {author} {\bibfnamefont {C.}~\bibnamefont {Benedetti}}, \bibinfo {author} {\bibfnamefont {E.}~\bibnamefont {Esarey}},\ and\ \bibinfo {author} {\bibfnamefont {W.}~\bibnamefont {Leemans}},\ }\bibfield  {title} {\bibinfo {title} {Laser-plasma-based linear collider using hollow plasma channels},\ }\href {https://doi.org/https://doi.org/10.1016/j.nima.2016.03.001} {\bibfield  {journal} {\bibinfo  {journal} {Nuclear Instruments and Methods in Physics Research Section A: Accelerators, Spectrometers, Detectors and Associated Equipment}\ }\textbf {\bibinfo {volume} {829}},\ \bibinfo {pages} {113} (\bibinfo {year} {2016})},\ \bibinfo {note} {2nd European Advanced Accelerator Concepts Workshop - EAAC 2015}\BibitemShut {NoStop}%
\bibitem [{\citenamefont {Tanaka}\ \emph {et~al.}(2020)\citenamefont {Tanaka}, \citenamefont {Spohr}, \citenamefont {Balabanski}, \citenamefont {Balascuta}, \citenamefont {Capponi}, \citenamefont {Cernaianu}, \citenamefont {Cuciuc}, \citenamefont {Cucoanes}, \citenamefont {Dancus}, \citenamefont {Dhal}, \citenamefont {Diaconescu}, \citenamefont {Doria}, \citenamefont {Ghenuche}, \citenamefont {Ghita}, \citenamefont {Kisyov}, \citenamefont {Nastasa}, \citenamefont {Ong}, \citenamefont {Rotaru}, \citenamefont {Sangwan}, \citenamefont {S{\"o}derstr{\"o}m}, \citenamefont {Stutman}, \citenamefont {Suliman}, \citenamefont {Tesileanu}, \citenamefont {Tudor}, \citenamefont {Tsoneva}, \citenamefont {Ur}, \citenamefont {Ursescu},\ and\ \citenamefont {Zamfir}}]{Tanaka2020cur}%
  \BibitemOpen
  \bibfield  {author} {\bibinfo {author} {\bibfnamefont {K.~A.}\ \bibnamefont {Tanaka}}, \bibinfo {author} {\bibfnamefont {K.~M.}\ \bibnamefont {Spohr}}, \bibinfo {author} {\bibfnamefont {D.~L.}\ \bibnamefont {Balabanski}}, \bibinfo {author} {\bibfnamefont {S.}~\bibnamefont {Balascuta}}, \bibinfo {author} {\bibfnamefont {L.}~\bibnamefont {Capponi}}, \bibinfo {author} {\bibfnamefont {M.~O.}\ \bibnamefont {Cernaianu}}, \bibinfo {author} {\bibfnamefont {M.}~\bibnamefont {Cuciuc}}, \bibinfo {author} {\bibfnamefont {A.}~\bibnamefont {Cucoanes}}, \bibinfo {author} {\bibfnamefont {I.}~\bibnamefont {Dancus}}, \bibinfo {author} {\bibfnamefont {A.}~\bibnamefont {Dhal}}, \bibinfo {author} {\bibfnamefont {B.}~\bibnamefont {Diaconescu}}, \bibinfo {author} {\bibfnamefont {D.}~\bibnamefont {Doria}}, \bibinfo {author} {\bibfnamefont {P.}~\bibnamefont {Ghenuche}}, \bibinfo {author} {\bibfnamefont {D.~G.}\ \bibnamefont {Ghita}}, \bibinfo {author} {\bibfnamefont {S.}~\bibnamefont {Kisyov}}, \bibinfo {author} {\bibfnamefont {V.}~\bibnamefont {Nastasa}}, \bibinfo {author} {\bibfnamefont {J.~F.}\ \bibnamefont {Ong}}, \bibinfo {author} {\bibfnamefont {F.}~\bibnamefont {Rotaru}}, \bibinfo {author} {\bibfnamefont {D.}~\bibnamefont {Sangwan}}, \bibinfo {author} {\bibfnamefont {P.-A.}\ \bibnamefont {S{\"o}derstr{\"o}m}}, \bibinfo {author} {\bibfnamefont {D.}~\bibnamefont {Stutman}}, \bibinfo {author} {\bibfnamefont {G.}~\bibnamefont {Suliman}}, \bibinfo {author} {\bibfnamefont {O.}~\bibnamefont {Tesileanu}}, \bibinfo {author} {\bibfnamefont {L.}~\bibnamefont {Tudor}}, \bibinfo {author} {\bibfnamefont {N.}~\bibnamefont {Tsoneva}}, \bibinfo {author} {\bibfnamefont {C.~A.}\ \bibnamefont {Ur}}, \bibinfo {author} {\bibfnamefont {D.}~\bibnamefont {Ursescu}},\ and\ \bibinfo {author} {\bibfnamefont {N.~V.}\ \bibnamefont {Zamfir}},\ }\bibfield  {title} {\bibinfo {title} {Current status and highlights of the {ELI-NP} research program},\ }\href {https://doi.org/10.1063/1.5093535} {\bibfield  {journal} {\bibinfo  {journal} {Matter and Radiation at
  Extremes}\ }\textbf {\bibinfo {volume} {5}},\ \bibinfo {pages} {024402} (\bibinfo {year} {2020})}\BibitemShut {NoStop}%
\bibitem [{\citenamefont {Gonsalves}\ \emph {et~al.}(2019)\citenamefont {Gonsalves}, \citenamefont {Nakamura}, \citenamefont {Daniels}, \citenamefont {Benedetti}, \citenamefont {Pieronek}, \citenamefont {de~Raadt}, \citenamefont {Steinke}, \citenamefont {Bin}, \citenamefont {Bulanov}, \citenamefont {van Tilborg}, \citenamefont {Geddes}, \citenamefont {Schroeder}, \citenamefont {T\'oth}, \citenamefont {Esarey}, \citenamefont {Swanson}, \citenamefont {Fan-Chiang}, \citenamefont {Bagdasarov}, \citenamefont {Bobrova}, \citenamefont {Gasilov}, \citenamefont {Korn}, \citenamefont {Sasorov},\ and\ \citenamefont {Leemans}}]{Gonsalves2019PW}%
  \BibitemOpen
  \bibfield  {author} {\bibinfo {author} {\bibfnamefont {A.~J.}\ \bibnamefont {Gonsalves}}, \bibinfo {author} {\bibfnamefont {K.}~\bibnamefont {Nakamura}}, \bibinfo {author} {\bibfnamefont {J.}~\bibnamefont {Daniels}}, \bibinfo {author} {\bibfnamefont {C.}~\bibnamefont {Benedetti}}, \bibinfo {author} {\bibfnamefont {C.}~\bibnamefont {Pieronek}}, \bibinfo {author} {\bibfnamefont {T.~C.~H.}\ \bibnamefont {de~Raadt}}, \bibinfo {author} {\bibfnamefont {S.}~\bibnamefont {Steinke}}, \bibinfo {author} {\bibfnamefont {J.~H.}\ \bibnamefont {Bin}}, \bibinfo {author} {\bibfnamefont {S.~S.}\ \bibnamefont {Bulanov}}, \bibinfo {author} {\bibfnamefont {J.}~\bibnamefont {van Tilborg}}, \bibinfo {author} {\bibfnamefont {C.~G.~R.}\ \bibnamefont {Geddes}}, \bibinfo {author} {\bibfnamefont {C.~B.}\ \bibnamefont {Schroeder}}, \bibinfo {author} {\bibfnamefont {C.}~\bibnamefont {T\'oth}}, \bibinfo {author} {\bibfnamefont {E.}~\bibnamefont {Esarey}}, \bibinfo {author} {\bibfnamefont {K.}~\bibnamefont {Swanson}}, \bibinfo {author} {\bibfnamefont {L.}~\bibnamefont {Fan-Chiang}}, \bibinfo {author} {\bibfnamefont {G.}~\bibnamefont {Bagdasarov}}, \bibinfo {author} {\bibfnamefont {N.}~\bibnamefont {Bobrova}}, \bibinfo {author} {\bibfnamefont {V.}~\bibnamefont {Gasilov}}, \bibinfo {author} {\bibfnamefont {G.}~\bibnamefont {Korn}}, \bibinfo {author} {\bibfnamefont {P.}~\bibnamefont {Sasorov}},\ and\ \bibinfo {author} {\bibfnamefont {W.~P.}\ \bibnamefont {Leemans}},\ }\bibfield  {title} {\bibinfo {title} {Petawatt laser guiding and electron beam acceleration to 8 {GeV} in a laser-heated capillary discharge waveguide},\ }\href {https://doi.org/10.1103/PhysRevLett.122.084801} {\bibfield  {journal} {\bibinfo  {journal} {Phys. Rev. Lett.}\ }\textbf {\bibinfo {volume} {122}},\ \bibinfo {pages} {084801} (\bibinfo {year} {2019})}\BibitemShut {NoStop}%
\bibitem [{\citenamefont {Kruer}(2019)}]{kruer2019physics}%
  \BibitemOpen
  \bibfield  {author} {\bibinfo {author} {\bibfnamefont {W.}~\bibnamefont {Kruer}},\ }\href@noop {} {\emph {\bibinfo {title} {The physics of laser plasma interactions}}}\ (\bibinfo  {publisher} {crc Press},\ \bibinfo {year} {2019})\BibitemShut {NoStop}%
\bibitem [{\citenamefont {Mora}\ and\ \citenamefont {Antonsen}(1997)}]{Mora1997kin}%
  \BibitemOpen
  \bibfield  {author} {\bibinfo {author} {\bibfnamefont {P.}~\bibnamefont {Mora}}\ and\ \bibinfo {author} {\bibfnamefont {T.~M.}\ \bibnamefont {Antonsen}, \bibfnamefont {Jr.}},\ }\bibfield  {title} {\bibinfo {title} {{Kinetic modeling of intense, short laser pulses propagating in tenuous plasmas}},\ }\href {https://doi.org/10.1063/1.872134} {\bibfield  {journal} {\bibinfo  {journal} {Physics of Plasmas}\ }\textbf {\bibinfo {volume} {4}},\ \bibinfo {pages} {217} (\bibinfo {year} {1997})}\BibitemShut {NoStop}%
\bibitem [{\citenamefont {Esarey}\ \emph {et~al.}(2009)\citenamefont {Esarey}, \citenamefont {Schroeder},\ and\ \citenamefont {Leemans}}]{Esarey2009PHY}%
  \BibitemOpen
  \bibfield  {author} {\bibinfo {author} {\bibfnamefont {E.}~\bibnamefont {Esarey}}, \bibinfo {author} {\bibfnamefont {C.~B.}\ \bibnamefont {Schroeder}},\ and\ \bibinfo {author} {\bibfnamefont {W.~P.}\ \bibnamefont {Leemans}},\ }\bibfield  {title} {\bibinfo {title} {Physics of laser-driven plasma-based electron accelerators},\ }\href {https://doi.org/10.1103/RevModPhys.81.1229} {\bibfield  {journal} {\bibinfo  {journal} {Rev. Mod. Phys.}\ }\textbf {\bibinfo {volume} {81}},\ \bibinfo {pages} {1229} (\bibinfo {year} {2009})}\BibitemShut {NoStop}%
\bibitem [{\citenamefont {Shukla}\ \emph {et~al.}(1986)\citenamefont {Shukla}, \citenamefont {Rao}, \citenamefont {Yu},\ and\ \citenamefont {Tsintsadze}}]{Shukla1986aa}%
  \BibitemOpen
  \bibfield  {author} {\bibinfo {author} {\bibfnamefont {P.~K.}\ \bibnamefont {Shukla}}, \bibinfo {author} {\bibfnamefont {N.~N.}\ \bibnamefont {Rao}}, \bibinfo {author} {\bibfnamefont {M.~Y.}\ \bibnamefont {Yu}},\ and\ \bibinfo {author} {\bibfnamefont {N.~L.}\ \bibnamefont {Tsintsadze}},\ }\bibfield  {title} {\bibinfo {title} {Relativistic nonlinear effects in plasmas},\ }\href {https://doi.org/https://doi.org/10.1016/0370-1573(86)90157-2} {\bibfield  {journal} {\bibinfo  {journal} {Physics Reports}\ }\textbf {\bibinfo {volume} {138}},\ \bibinfo {pages} {1} (\bibinfo {year} {1986})}\BibitemShut {NoStop}%
\bibitem [{\citenamefont {Terzani}\ \emph {et~al.}(2021)\citenamefont {Terzani}, \citenamefont {Benedetti}, \citenamefont {Schroeder},\ and\ \citenamefont {Esarey}}]{Terzani2021acc}%
  \BibitemOpen
  \bibfield  {author} {\bibinfo {author} {\bibfnamefont {D.}~\bibnamefont {Terzani}}, \bibinfo {author} {\bibfnamefont {C.}~\bibnamefont {Benedetti}}, \bibinfo {author} {\bibfnamefont {C.~B.}\ \bibnamefont {Schroeder}},\ and\ \bibinfo {author} {\bibfnamefont {E.}~\bibnamefont {Esarey}},\ }\bibfield  {title} {\bibinfo {title} {{Accuracy of the time-averaged ponderomotive approximation for laser-plasma accelerator modeling}},\ }\href {https://doi.org/10.1063/5.0050580} {\bibfield  {journal} {\bibinfo  {journal} {Physics of Plasmas}\ }\textbf {\bibinfo {volume} {28}},\ \bibinfo {pages} {063105} (\bibinfo {year} {2021})}\BibitemShut {NoStop}%
\bibitem [{\citenamefont {Tuev}\ \emph {et~al.}(2023)\citenamefont {Tuev}, \citenamefont {Spitsyn},\ and\ \citenamefont {Lotov}}]{Tuev:2023aa}%
  \BibitemOpen
  \bibfield  {author} {\bibinfo {author} {\bibfnamefont {P.~V.}\ \bibnamefont {Tuev}}, \bibinfo {author} {\bibfnamefont {R.~I.}\ \bibnamefont {Spitsyn}},\ and\ \bibinfo {author} {\bibfnamefont {K.~V.}\ \bibnamefont {Lotov}},\ }\bibfield  {title} {\bibinfo {title} {Advanced quasistatic approximation},\ }\href {https://doi.org/10.1134/S1063780X22601249} {\bibfield  {journal} {\bibinfo  {journal} {Plasma Physics Reports}\ }\textbf {\bibinfo {volume} {49}},\ \bibinfo {pages} {229} (\bibinfo {year} {2023})}\BibitemShut {NoStop}%
\bibitem [{\citenamefont {Ferri}\ \emph {et~al.}(2016)\citenamefont {Ferri}, \citenamefont {Davoine}, \citenamefont {Fourmaux}, \citenamefont {Kieffer}, \citenamefont {Corde}, \citenamefont {Ta~Phuoc},\ and\ \citenamefont {Lifschitz}}]{Ferri2016aa}%
  \BibitemOpen
  \bibfield  {author} {\bibinfo {author} {\bibfnamefont {J.}~\bibnamefont {Ferri}}, \bibinfo {author} {\bibfnamefont {X.}~\bibnamefont {Davoine}}, \bibinfo {author} {\bibfnamefont {S.}~\bibnamefont {Fourmaux}}, \bibinfo {author} {\bibfnamefont {J.~C.}\ \bibnamefont {Kieffer}}, \bibinfo {author} {\bibfnamefont {S.}~\bibnamefont {Corde}}, \bibinfo {author} {\bibfnamefont {K.}~\bibnamefont {Ta~Phuoc}},\ and\ \bibinfo {author} {\bibfnamefont {A.}~\bibnamefont {Lifschitz}},\ }\bibfield  {title} {\bibinfo {title} {Effect of experimental laser imperfections on laser wakefield acceleration and betatron source},\ }\href {https://doi.org/10.1038/srep27846} {\bibfield  {journal} {\bibinfo  {journal} {Scientific Reports}\ }\textbf {\bibinfo {volume} {6}},\ \bibinfo {pages} {27846} (\bibinfo {year} {2016})}\BibitemShut {NoStop}%
\bibitem [{\citenamefont {Corde}\ \emph {et~al.}(2013)\citenamefont {Corde}, \citenamefont {Thaury}, \citenamefont {Lifschitz}, \citenamefont {Lambert}, \citenamefont {Ta~Phuoc}, \citenamefont {Davoine}, \citenamefont {Lehe}, \citenamefont {Douillet}, \citenamefont {Rousse},\ and\ \citenamefont {Malka}}]{Corde2013aa}%
  \BibitemOpen
  \bibfield  {author} {\bibinfo {author} {\bibfnamefont {S.}~\bibnamefont {Corde}}, \bibinfo {author} {\bibfnamefont {C.}~\bibnamefont {Thaury}}, \bibinfo {author} {\bibfnamefont {A.}~\bibnamefont {Lifschitz}}, \bibinfo {author} {\bibfnamefont {G.}~\bibnamefont {Lambert}}, \bibinfo {author} {\bibfnamefont {K.}~\bibnamefont {Ta~Phuoc}}, \bibinfo {author} {\bibfnamefont {X.}~\bibnamefont {Davoine}}, \bibinfo {author} {\bibfnamefont {R.}~\bibnamefont {Lehe}}, \bibinfo {author} {\bibfnamefont {D.}~\bibnamefont {Douillet}}, \bibinfo {author} {\bibfnamefont {A.}~\bibnamefont {Rousse}},\ and\ \bibinfo {author} {\bibfnamefont {V.}~\bibnamefont {Malka}},\ }\bibfield  {title} {\bibinfo {title} {Observation of longitudinal and transverse self-injections in laser-plasma accelerators},\ }\href {https://doi.org/10.1038/ncomms2528} {\bibfield  {journal} {\bibinfo  {journal} {Nature Communications}\ }\textbf {\bibinfo {volume} {4}},\ \bibinfo {pages} {1501} (\bibinfo {year} {2013})}\BibitemShut {NoStop}%
\bibitem [{\citenamefont {Nerush}\ and\ \citenamefont {Kostyukov}(2009)}]{Nerush2009carr}%
  \BibitemOpen
  \bibfield  {author} {\bibinfo {author} {\bibfnamefont {E.~N.}\ \bibnamefont {Nerush}}\ and\ \bibinfo {author} {\bibfnamefont {I.~Y.}\ \bibnamefont {Kostyukov}},\ }\bibfield  {title} {\bibinfo {title} {Carrier-envelope phase effects in plasma-based electron acceleration with few-cycle laser pulses},\ }\href {https://doi.org/10.1103/PhysRevLett.103.035001} {\bibfield  {journal} {\bibinfo  {journal} {Phys. Rev. Lett.}\ }\textbf {\bibinfo {volume} {103}},\ \bibinfo {pages} {035001} (\bibinfo {year} {2009})}\BibitemShut {NoStop}%
\bibitem [{\citenamefont {Seidel}\ \emph {et~al.}(2024)\citenamefont {Seidel}, \citenamefont {Lei}, \citenamefont {Zepter}, \citenamefont {Kaluza}, \citenamefont {S\"avert}, \citenamefont {Zepf},\ and\ \citenamefont {Seipt}}]{Seidel2022pointing}%
  \BibitemOpen
  \bibfield  {author} {\bibinfo {author} {\bibfnamefont {A.}~\bibnamefont {Seidel}}, \bibinfo {author} {\bibfnamefont {B.}~\bibnamefont {Lei}}, \bibinfo {author} {\bibfnamefont {C.}~\bibnamefont {Zepter}}, \bibinfo {author} {\bibfnamefont {M.~C.}\ \bibnamefont {Kaluza}}, \bibinfo {author} {\bibfnamefont {A.}~\bibnamefont {S\"avert}}, \bibinfo {author} {\bibfnamefont {M.}~\bibnamefont {Zepf}},\ and\ \bibinfo {author} {\bibfnamefont {D.}~\bibnamefont {Seipt}},\ }\bibfield  {title} {\bibinfo {title} {Polarization and cep dependence of the transverse phase space in laser driven accelerators},\ }\href {https://doi.org/10.1103/PhysRevResearch.6.013056} {\bibfield  {journal} {\bibinfo  {journal} {Phys. Rev. Res.}\ }\textbf {\bibinfo {volume} {6}},\ \bibinfo {pages} {013056} (\bibinfo {year} {2024})}\BibitemShut {NoStop}%
\bibitem [{\citenamefont {Pollock}\ \emph {et~al.}(2015)\citenamefont {Pollock}, \citenamefont {Tsung}, \citenamefont {Albert}, \citenamefont {Shaw}, \citenamefont {Clayton}, \citenamefont {Davidson}, \citenamefont {Lemos}, \citenamefont {Marsh}, \citenamefont {Pak}, \citenamefont {Ralph}, \citenamefont {Mori},\ and\ \citenamefont {Joshi}}]{Pollock2015for}%
  \BibitemOpen
  \bibfield  {author} {\bibinfo {author} {\bibfnamefont {B.~B.}\ \bibnamefont {Pollock}}, \bibinfo {author} {\bibfnamefont {F.~S.}\ \bibnamefont {Tsung}}, \bibinfo {author} {\bibfnamefont {F.}~\bibnamefont {Albert}}, \bibinfo {author} {\bibfnamefont {J.~L.}\ \bibnamefont {Shaw}}, \bibinfo {author} {\bibfnamefont {C.~E.}\ \bibnamefont {Clayton}}, \bibinfo {author} {\bibfnamefont {A.}~\bibnamefont {Davidson}}, \bibinfo {author} {\bibfnamefont {N.}~\bibnamefont {Lemos}}, \bibinfo {author} {\bibfnamefont {K.~A.}\ \bibnamefont {Marsh}}, \bibinfo {author} {\bibfnamefont {A.}~\bibnamefont {Pak}}, \bibinfo {author} {\bibfnamefont {J.~E.}\ \bibnamefont {Ralph}}, \bibinfo {author} {\bibfnamefont {W.~B.}\ \bibnamefont {Mori}},\ and\ \bibinfo {author} {\bibfnamefont {C.}~\bibnamefont {Joshi}},\ }\bibfield  {title} {\bibinfo {title} {Formation of ultrarelativistic electron rings from a laser-wakefield accelerator},\ }\href {https://doi.org/10.1103/PhysRevLett.115.055004} {\bibfield  {journal} {\bibinfo  {journal} {Phys. Rev. Lett.}\ }\textbf {\bibinfo {volume} {115}},\ \bibinfo {pages} {055004} (\bibinfo {year} {2015})}\BibitemShut {NoStop}%
\bibitem [{\citenamefont {Ma}\ \emph {et~al.}(2016)\citenamefont {Ma}, \citenamefont {Chen}, \citenamefont {Li}, \citenamefont {Yan}, \citenamefont {Huang}, \citenamefont {Chen}, \citenamefont {Sheng}, \citenamefont {Nakajima}, \citenamefont {Tajima},\ and\ \citenamefont {Zhang}}]{Ma2016aa}%
  \BibitemOpen
  \bibfield  {author} {\bibinfo {author} {\bibfnamefont {Y.}~\bibnamefont {Ma}}, \bibinfo {author} {\bibfnamefont {L.}~\bibnamefont {Chen}}, \bibinfo {author} {\bibfnamefont {D.}~\bibnamefont {Li}}, \bibinfo {author} {\bibfnamefont {W.}~\bibnamefont {Yan}}, \bibinfo {author} {\bibfnamefont {K.}~\bibnamefont {Huang}}, \bibinfo {author} {\bibfnamefont {M.}~\bibnamefont {Chen}}, \bibinfo {author} {\bibfnamefont {Z.}~\bibnamefont {Sheng}}, \bibinfo {author} {\bibfnamefont {K.}~\bibnamefont {Nakajima}}, \bibinfo {author} {\bibfnamefont {T.}~\bibnamefont {Tajima}},\ and\ \bibinfo {author} {\bibfnamefont {J.}~\bibnamefont {Zhang}},\ }\bibfield  {title} {\bibinfo {title} {Generation of femtosecond $\gamma$-ray bursts stimulated by laser-driven hosing evolution-ray bursts stimulated by laser-driven hosing evolution},\ }\href {https://doi.org/10.1038/srep30491} {\bibfield  {journal} {\bibinfo  {journal} {Scientific Reports}\ }\textbf {\bibinfo {volume} {6}},\ \bibinfo {pages} {30491} (\bibinfo {year} {2016})}\BibitemShut {NoStop}%
\bibitem [{\citenamefont {Satanin}\ \emph {et~al.}(2014)\citenamefont {Satanin}, \citenamefont {Denisenko}, \citenamefont {Gelman},\ and\ \citenamefont {Nori}}]{Satanin2014amp}%
  \BibitemOpen
  \bibfield  {author} {\bibinfo {author} {\bibfnamefont {A.~M.}\ \bibnamefont {Satanin}}, \bibinfo {author} {\bibfnamefont {M.~V.}\ \bibnamefont {Denisenko}}, \bibinfo {author} {\bibfnamefont {A.~I.}\ \bibnamefont {Gelman}},\ and\ \bibinfo {author} {\bibfnamefont {F.}~\bibnamefont {Nori}},\ }\bibfield  {title} {\bibinfo {title} {Amplitude and phase effects in josephson qubits driven by a biharmonic electromagnetic field},\ }\href {https://doi.org/10.1103/PhysRevB.90.104516} {\bibfield  {journal} {\bibinfo  {journal} {Phys. Rev. B}\ }\textbf {\bibinfo {volume} {90}},\ \bibinfo {pages} {104516} (\bibinfo {year} {2014})}\BibitemShut {NoStop}%
\bibitem [{\citenamefont {Kim}\ \emph {et~al.}(2021)\citenamefont {Kim}, \citenamefont {Wang}, \citenamefont {Khudik},\ and\ \citenamefont {Shvets}}]{Kim2021sub}%
  \BibitemOpen
  \bibfield  {author} {\bibinfo {author} {\bibfnamefont {J.}~\bibnamefont {Kim}}, \bibinfo {author} {\bibfnamefont {T.}~\bibnamefont {Wang}}, \bibinfo {author} {\bibfnamefont {V.}~\bibnamefont {Khudik}},\ and\ \bibinfo {author} {\bibfnamefont {G.}~\bibnamefont {Shvets}},\ }\bibfield  {title} {\bibinfo {title} {Subfemtosecond wakefield injector and accelerator based on an undulating plasma bubble controlled by a laser phase},\ }\href {https://doi.org/10.1103/PhysRevLett.127.164801} {\bibfield  {journal} {\bibinfo  {journal} {Phys. Rev. Lett.}\ }\textbf {\bibinfo {volume} {127}},\ \bibinfo {pages} {164801} (\bibinfo {year} {2021})}\BibitemShut {NoStop}%
\bibitem [{\citenamefont {Huijts}\ \emph {et~al.}(2022)\citenamefont {Huijts}, \citenamefont {Rovige}, \citenamefont {Andriyash}, \citenamefont {Vernier}, \citenamefont {Ouill\'e}, \citenamefont {Kaur}, \citenamefont {Cheng}, \citenamefont {Lopez-Martens},\ and\ \citenamefont {Faure}}]{Huijts2022wave}%
  \BibitemOpen
  \bibfield  {author} {\bibinfo {author} {\bibfnamefont {J.}~\bibnamefont {Huijts}}, \bibinfo {author} {\bibfnamefont {L.}~\bibnamefont {Rovige}}, \bibinfo {author} {\bibfnamefont {I.~A.}\ \bibnamefont {Andriyash}}, \bibinfo {author} {\bibfnamefont {A.}~\bibnamefont {Vernier}}, \bibinfo {author} {\bibfnamefont {M.}~\bibnamefont {Ouill\'e}}, \bibinfo {author} {\bibfnamefont {J.}~\bibnamefont {Kaur}}, \bibinfo {author} {\bibfnamefont {Z.}~\bibnamefont {Cheng}}, \bibinfo {author} {\bibfnamefont {R.}~\bibnamefont {Lopez-Martens}},\ and\ \bibinfo {author} {\bibfnamefont {J.}~\bibnamefont {Faure}},\ }\bibfield  {title} {\bibinfo {title} {Waveform control of relativistic electron dynamics in laser-plasma acceleration},\ }\href {https://doi.org/10.1103/PhysRevX.12.011036} {\bibfield  {journal} {\bibinfo  {journal} {Phys. Rev. X}\ }\textbf {\bibinfo {volume} {12}},\ \bibinfo {pages} {011036} (\bibinfo {year} {2022})}\BibitemShut {NoStop}%
\bibitem [{\citenamefont {Chen}\ \emph {et~al.}(2021)\citenamefont {Chen}, \citenamefont {Xu}, \citenamefont {Tang}, \citenamefont {Wang},\ and\ \citenamefont {Li}}]{Chen2021enh}%
  \BibitemOpen
  \bibfield  {author} {\bibinfo {author} {\bibfnamefont {J.}~\bibnamefont {Chen}}, \bibinfo {author} {\bibfnamefont {S.}~\bibnamefont {Xu}}, \bibinfo {author} {\bibfnamefont {N.}~\bibnamefont {Tang}}, \bibinfo {author} {\bibfnamefont {S.}~\bibnamefont {Wang}},\ and\ \bibinfo {author} {\bibfnamefont {Z.}~\bibnamefont {Li}},\ }\bibfield  {title} {\bibinfo {title} {Enhanced soft x-ray betatron radiation from a transversely oscillating laser plasma wake},\ }\href {https://doi.org/10.1364/OE.420150} {\bibfield  {journal} {\bibinfo  {journal} {Opt. Express}\ }\textbf {\bibinfo {volume} {29}},\ \bibinfo {pages} {13302} (\bibinfo {year} {2021})}\BibitemShut {NoStop}%
\bibitem [{\citenamefont {Rakowski}\ \emph {et~al.}(2022)\citenamefont {Rakowski}, \citenamefont {Zhang}, \citenamefont {Jensen}, \citenamefont {Kettle}, \citenamefont {Kawamoto}, \citenamefont {Banerjee}, \citenamefont {Fruhling}, \citenamefont {Golovin}, \citenamefont {Haden}, \citenamefont {Robinson}, \citenamefont {Umstadter}, \citenamefont {Shadwick},\ and\ \citenamefont {Fuchs}}]{Rakowski2022aa}%
  \BibitemOpen
  \bibfield  {author} {\bibinfo {author} {\bibfnamefont {R.}~\bibnamefont {Rakowski}}, \bibinfo {author} {\bibfnamefont {P.}~\bibnamefont {Zhang}}, \bibinfo {author} {\bibfnamefont {K.}~\bibnamefont {Jensen}}, \bibinfo {author} {\bibfnamefont {B.}~\bibnamefont {Kettle}}, \bibinfo {author} {\bibfnamefont {T.}~\bibnamefont {Kawamoto}}, \bibinfo {author} {\bibfnamefont {S.}~\bibnamefont {Banerjee}}, \bibinfo {author} {\bibfnamefont {C.}~\bibnamefont {Fruhling}}, \bibinfo {author} {\bibfnamefont {G.}~\bibnamefont {Golovin}}, \bibinfo {author} {\bibfnamefont {D.}~\bibnamefont {Haden}}, \bibinfo {author} {\bibfnamefont {M.~S.}\ \bibnamefont {Robinson}}, \bibinfo {author} {\bibfnamefont {D.}~\bibnamefont {Umstadter}}, \bibinfo {author} {\bibfnamefont {B.~A.}\ \bibnamefont {Shadwick}},\ and\ \bibinfo {author} {\bibfnamefont {M.}~\bibnamefont {Fuchs}},\ }\bibfield  {title} {\bibinfo {title} {Transverse oscillating bubble enhanced laser-driven betatron x-ray radiation generation},\ }\href {https://doi.org/10.1038/s41598-022-14748-z} {\bibfield  {journal} {\bibinfo  {journal} {Scientific Reports}\ }\textbf {\bibinfo {volume} {12}},\ \bibinfo {pages} {10855} (\bibinfo {year} {2022})}\BibitemShut {NoStop}%
\bibitem [{\citenamefont {Mishra}\ \emph {et~al.}(2022)\citenamefont {Mishra}, \citenamefont {Rao}, \citenamefont {Moorti},\ and\ \citenamefont {Chakera}}]{Mishra2022enh}%
  \BibitemOpen
  \bibfield  {author} {\bibinfo {author} {\bibfnamefont {S.}~\bibnamefont {Mishra}}, \bibinfo {author} {\bibfnamefont {B.~S.}\ \bibnamefont {Rao}}, \bibinfo {author} {\bibfnamefont {A.}~\bibnamefont {Moorti}},\ and\ \bibinfo {author} {\bibfnamefont {J.~A.}\ \bibnamefont {Chakera}},\ }\bibfield  {title} {\bibinfo {title} {Enhanced betatron x-ray emission in a laser wakefield accelerator and wiggler due to collective oscillations of electrons},\ }\href {https://doi.org/10.1103/PhysRevAccelBeams.25.090703} {\bibfield  {journal} {\bibinfo  {journal} {Phys. Rev. Accel. Beams}\ }\textbf {\bibinfo {volume} {25}},\ \bibinfo {pages} {090703} (\bibinfo {year} {2022})}\BibitemShut {NoStop}%
\bibitem [{\citenamefont {Ma}\ \emph {et~al.}(2020)\citenamefont {Ma}, \citenamefont {Seipt}, \citenamefont {Hussein}, \citenamefont {Hakimi}, \citenamefont {Beier}, \citenamefont {Hansen}, \citenamefont {Hinojosa}, \citenamefont {Maksimchuk}, \citenamefont {Nees}, \citenamefont {Krushelnick}, \citenamefont {Thomas},\ and\ \citenamefont {Dollar}}]{Ma2020pol}%
  \BibitemOpen
  \bibfield  {author} {\bibinfo {author} {\bibfnamefont {Y.}~\bibnamefont {Ma}}, \bibinfo {author} {\bibfnamefont {D.}~\bibnamefont {Seipt}}, \bibinfo {author} {\bibfnamefont {A.~E.}\ \bibnamefont {Hussein}}, \bibinfo {author} {\bibfnamefont {S.}~\bibnamefont {Hakimi}}, \bibinfo {author} {\bibfnamefont {N.~F.}\ \bibnamefont {Beier}}, \bibinfo {author} {\bibfnamefont {S.~B.}\ \bibnamefont {Hansen}}, \bibinfo {author} {\bibfnamefont {J.}~\bibnamefont {Hinojosa}}, \bibinfo {author} {\bibfnamefont {A.}~\bibnamefont {Maksimchuk}}, \bibinfo {author} {\bibfnamefont {J.}~\bibnamefont {Nees}}, \bibinfo {author} {\bibfnamefont {K.}~\bibnamefont {Krushelnick}}, \bibinfo {author} {\bibfnamefont {A.~G.~R.}\ \bibnamefont {Thomas}},\ and\ \bibinfo {author} {\bibfnamefont {F.}~\bibnamefont {Dollar}},\ }\bibfield  {title} {\bibinfo {title} {Polarization-dependent self-injection by above threshold ionization heating in a laser wakefield accelerator},\ }\href {https://doi.org/10.1103/PhysRevLett.124.114801} {\bibfield  {journal} {\bibinfo  {journal} {Phys. Rev. Lett.}\ }\textbf {\bibinfo {volume} {124}},\ \bibinfo {pages} {114801} (\bibinfo {year} {2020})}\BibitemShut {NoStop}%
\bibitem [{\citenamefont {Gallardo~Gonz{\'a}lez}\ \emph {et~al.}(2018)\citenamefont {Gallardo~Gonz{\'a}lez}, \citenamefont {Ekerfelt}, \citenamefont {Hansson}, \citenamefont {Audet}, \citenamefont {Aurand}, \citenamefont {Desforges}, \citenamefont {Dufr{\'e}noy}, \citenamefont {Persson}, \citenamefont {Davoine}, \citenamefont {Wahlstr{\"o}m}, \citenamefont {Cros},\ and\ \citenamefont {Lundh}}]{GallardoGonzalez2018aa}%
  \BibitemOpen
  \bibfield  {author} {\bibinfo {author} {\bibfnamefont {I.}~\bibnamefont {Gallardo~Gonz{\'a}lez}}, \bibinfo {author} {\bibfnamefont {H.}~\bibnamefont {Ekerfelt}}, \bibinfo {author} {\bibfnamefont {M.}~\bibnamefont {Hansson}}, \bibinfo {author} {\bibfnamefont {T.~L.}\ \bibnamefont {Audet}}, \bibinfo {author} {\bibfnamefont {B.}~\bibnamefont {Aurand}}, \bibinfo {author} {\bibfnamefont {F.~G.}\ \bibnamefont {Desforges}}, \bibinfo {author} {\bibfnamefont {S.~D.}\ \bibnamefont {Dufr{\'e}noy}}, \bibinfo {author} {\bibfnamefont {A.}~\bibnamefont {Persson}}, \bibinfo {author} {\bibfnamefont {X.}~\bibnamefont {Davoine}}, \bibinfo {author} {\bibfnamefont {C.-G.}\ \bibnamefont {Wahlstr{\"o}m}}, \bibinfo {author} {\bibfnamefont {B.}~\bibnamefont {Cros}},\ and\ \bibinfo {author} {\bibfnamefont {O.}~\bibnamefont {Lundh}},\ }\bibfield  {title} {\bibinfo {title} {Effects of the dopant concentration in laser wakefield and direct laser acceleration of electrons},\ }\href {https://doi.org/10.1088/1367-2630/aabe14} {\bibfield  {journal} {\bibinfo  {journal} {New Journal of Physics}\ }\textbf {\bibinfo {volume} {20}},\ \bibinfo {pages} {053011} (\bibinfo {year} {2018})}\BibitemShut {NoStop}%
\bibitem [{\citenamefont {Feng}\ \emph {et~al.}(2020)\citenamefont {Feng}, \citenamefont {Li}, \citenamefont {Geng}, \citenamefont {Li}, \citenamefont {Wang}, \citenamefont {Mirzaie},\ and\ \citenamefont {Chen}}]{Feng2020aa}%
  \BibitemOpen
  \bibfield  {author} {\bibinfo {author} {\bibfnamefont {J.}~\bibnamefont {Feng}}, \bibinfo {author} {\bibfnamefont {Y.}~\bibnamefont {Li}}, \bibinfo {author} {\bibfnamefont {X.}~\bibnamefont {Geng}}, \bibinfo {author} {\bibfnamefont {D.}~\bibnamefont {Li}}, \bibinfo {author} {\bibfnamefont {J.}~\bibnamefont {Wang}}, \bibinfo {author} {\bibfnamefont {M.}~\bibnamefont {Mirzaie}},\ and\ \bibinfo {author} {\bibfnamefont {L.}~\bibnamefont {Chen}},\ }\bibfield  {title} {\bibinfo {title} {Circularly polarized x-ray generation from an ionization induced laser plasma electron accelerator},\ }\href {https://doi.org/10.1088/1361-6587/abaf0b} {\bibfield  {journal} {\bibinfo  {journal} {Plasma Physics and Controlled Fusion}\ }\textbf {\bibinfo {volume} {62}},\ \bibinfo {pages} {105021} (\bibinfo {year} {2020})}\BibitemShut {NoStop}%
\bibitem [{\citenamefont {Lu}\ \emph {et~al.}(2007)\citenamefont {Lu}, \citenamefont {Tzoufras}, \citenamefont {Joshi}, \citenamefont {Tsung}, \citenamefont {Mori}, \citenamefont {Vieira}, \citenamefont {Fonseca},\ and\ \citenamefont {Silva}}]{Lu2007gen}%
  \BibitemOpen
  \bibfield  {author} {\bibinfo {author} {\bibfnamefont {W.}~\bibnamefont {Lu}}, \bibinfo {author} {\bibfnamefont {M.}~\bibnamefont {Tzoufras}}, \bibinfo {author} {\bibfnamefont {C.}~\bibnamefont {Joshi}}, \bibinfo {author} {\bibfnamefont {F.~S.}\ \bibnamefont {Tsung}}, \bibinfo {author} {\bibfnamefont {W.~B.}\ \bibnamefont {Mori}}, \bibinfo {author} {\bibfnamefont {J.}~\bibnamefont {Vieira}}, \bibinfo {author} {\bibfnamefont {R.~A.}\ \bibnamefont {Fonseca}},\ and\ \bibinfo {author} {\bibfnamefont {L.~O.}\ \bibnamefont {Silva}},\ }\bibfield  {title} {\bibinfo {title} {Generating multi-{GeV} electron bunches using single stage laser wakefield acceleration in a {3D} nonlinear regime},\ }\href {https://doi.org/10.1103/PhysRevSTAB.10.061301} {\bibfield  {journal} {\bibinfo  {journal} {Phys. Rev. ST Accel. Beams}\ }\textbf {\bibinfo {volume} {10}},\ \bibinfo {pages} {061301} (\bibinfo {year} {2007})}\BibitemShut {NoStop}%
\bibitem [{\citenamefont {Derouillat}\ \emph {et~al.}(2018)\citenamefont {Derouillat}, \citenamefont {Beck}, \citenamefont {P{\'e}rez}, \citenamefont {Vinci}, \citenamefont {Chiaramello}, \citenamefont {Grassi}, \citenamefont {Fl{\'e}}, \citenamefont {Bouchard}, \citenamefont {Plotnikov}, \citenamefont {Aunai}, \citenamefont {Dargent}, \citenamefont {Riconda},\ and\ \citenamefont {Grech}}]{derouillat_smilei_2018}%
  \BibitemOpen
  \bibfield  {author} {\bibinfo {author} {\bibfnamefont {J.}~\bibnamefont {Derouillat}}, \bibinfo {author} {\bibfnamefont {A.}~\bibnamefont {Beck}}, \bibinfo {author} {\bibfnamefont {F.}~\bibnamefont {P{\'e}rez}}, \bibinfo {author} {\bibfnamefont {T.}~\bibnamefont {Vinci}}, \bibinfo {author} {\bibfnamefont {M.}~\bibnamefont {Chiaramello}}, \bibinfo {author} {\bibfnamefont {A.}~\bibnamefont {Grassi}}, \bibinfo {author} {\bibfnamefont {M.}~\bibnamefont {Fl{\'e}}}, \bibinfo {author} {\bibfnamefont {G.}~\bibnamefont {Bouchard}}, \bibinfo {author} {\bibfnamefont {I.}~\bibnamefont {Plotnikov}}, \bibinfo {author} {\bibfnamefont {N.}~\bibnamefont {Aunai}}, \bibinfo {author} {\bibfnamefont {J.}~\bibnamefont {Dargent}}, \bibinfo {author} {\bibfnamefont {C.}~\bibnamefont {Riconda}},\ and\ \bibinfo {author} {\bibfnamefont {M.}~\bibnamefont {Grech}},\ }\bibfield  {title} {\bibinfo {title} {Smilei : {A} collaborative, open-source, multi-purpose particle-in-cell code for plasma simulation},\ }\href {https://doi.org/10.1016/j.cpc.2017.09.024} {\bibfield  {journal} {\bibinfo  {journal} {Computer Physics Communications}\ }\textbf {\bibinfo {volume} {222}},\ \bibinfo {pages} {351} (\bibinfo {year} {2018})}\BibitemShut {NoStop}%
\bibitem [{\citenamefont {Zhang}\ \emph {et~al.}(2012)\citenamefont {Zhang}, \citenamefont {Shen}, \citenamefont {Ji}, \citenamefont {Wang}, \citenamefont {Xu}, \citenamefont {Yu}, \citenamefont {Yi}, \citenamefont {Wang}, \citenamefont {Hafz},\ and\ \citenamefont {Kulagin}}]{Zhang2012eff}%
  \BibitemOpen
  \bibfield  {author} {\bibinfo {author} {\bibfnamefont {X.}~\bibnamefont {Zhang}}, \bibinfo {author} {\bibfnamefont {B.}~\bibnamefont {Shen}}, \bibinfo {author} {\bibfnamefont {L.}~\bibnamefont {Ji}}, \bibinfo {author} {\bibfnamefont {W.}~\bibnamefont {Wang}}, \bibinfo {author} {\bibfnamefont {J.}~\bibnamefont {Xu}}, \bibinfo {author} {\bibfnamefont {Y.}~\bibnamefont {Yu}}, \bibinfo {author} {\bibfnamefont {L.}~\bibnamefont {Yi}}, \bibinfo {author} {\bibfnamefont {X.}~\bibnamefont {Wang}}, \bibinfo {author} {\bibfnamefont {N.~A.~M.}\ \bibnamefont {Hafz}},\ and\ \bibinfo {author} {\bibfnamefont {V.}~\bibnamefont {Kulagin}},\ }\bibfield  {title} {\bibinfo {title} {{Effect of pulse profile and chirp on a laser wakefield generation}},\ }\href {https://doi.org/10.1063/1.4714610} {\bibfield  {journal} {\bibinfo  {journal} {Physics of Plasmas}\ }\textbf {\bibinfo {volume} {19}},\ \bibinfo {pages} {053103} (\bibinfo {year} {2012})}\BibitemShut {NoStop}%
\bibitem [{\citenamefont {Pathak}\ \emph {et~al.}(2012)\citenamefont {Pathak}, \citenamefont {Vieira}, \citenamefont {Fonseca},\ and\ \citenamefont {Silva}}]{Pathak2012eff}%
  \BibitemOpen
  \bibfield  {author} {\bibinfo {author} {\bibfnamefont {V.~B.}\ \bibnamefont {Pathak}}, \bibinfo {author} {\bibfnamefont {J.}~\bibnamefont {Vieira}}, \bibinfo {author} {\bibfnamefont {R.~A.}\ \bibnamefont {Fonseca}},\ and\ \bibinfo {author} {\bibfnamefont {L.~O.}\ \bibnamefont {Silva}},\ }\bibfield  {title} {\bibinfo {title} {Effect of the frequency chirp on laser wakefield acceleration},\ }\href {https://doi.org/10.1088/1367-2630/14/2/023057} {\bibfield  {journal} {\bibinfo  {journal} {New Journal of Physics}\ }\textbf {\bibinfo {volume} {14}},\ \bibinfo {pages} {023057} (\bibinfo {year} {2012})}\BibitemShut {NoStop}%
\bibitem [{\citenamefont {Sohbatzadeh}\ and\ \citenamefont {Akou}(2013)}]{Sohbatzadeh2013gro}%
  \BibitemOpen
  \bibfield  {author} {\bibinfo {author} {\bibfnamefont {F.}~\bibnamefont {Sohbatzadeh}}\ and\ \bibinfo {author} {\bibfnamefont {H.}~\bibnamefont {Akou}},\ }\bibfield  {title} {\bibinfo {title} {{Group velocity dispersion and relativistic effects on the wakefield induced by chirped laser pulse in parabolic plasma channel}},\ }\href {https://doi.org/10.1063/1.4798530} {\bibfield  {journal} {\bibinfo  {journal} {Physics of Plasmas}\ }\textbf {\bibinfo {volume} {20}},\ \bibinfo {pages} {043101} (\bibinfo {year} {2013})}\BibitemShut {NoStop}%
\bibitem [{\citenamefont {Huijts}\ \emph {et~al.}(2021)\citenamefont {Huijts}, \citenamefont {Andriyash}, \citenamefont {Rovige}, \citenamefont {Vernier},\ and\ \citenamefont {Faure}}]{Huijts2021ide}%
  \BibitemOpen
  \bibfield  {author} {\bibinfo {author} {\bibfnamefont {J.}~\bibnamefont {Huijts}}, \bibinfo {author} {\bibfnamefont {I.~A.}\ \bibnamefont {Andriyash}}, \bibinfo {author} {\bibfnamefont {L.}~\bibnamefont {Rovige}}, \bibinfo {author} {\bibfnamefont {A.}~\bibnamefont {Vernier}},\ and\ \bibinfo {author} {\bibfnamefont {J.}~\bibnamefont {Faure}},\ }\bibfield  {title} {\bibinfo {title} {Identifying observable carrier-envelope phase effects in laser wakefield acceleration with near-single-cycle pulses},\ }\href {https://doi.org/10.1063/5.0037925} {\bibfield  {journal} {\bibinfo  {journal} {Physics of Plasmas}\ }\textbf {\bibinfo {volume} {28}},\ \bibinfo {pages} {043101} (\bibinfo {year} {2021})}\BibitemShut {NoStop}%
\bibitem [{\citenamefont {Liu}\ \emph {et~al.}(2023)\citenamefont {Liu}, \citenamefont {Zhang}, \citenamefont {Yang}, \citenamefont {Ma}, \citenamefont {Cui}, \citenamefont {Li}, \citenamefont {Zou}, \citenamefont {Du}, \citenamefont {Zhao}, \citenamefont {Wang},\ and\ \citenamefont {Shao}}]{Liu2023bub}%
  \BibitemOpen
  \bibfield  {author} {\bibinfo {author} {\bibfnamefont {S.}~\bibnamefont {Liu}}, \bibinfo {author} {\bibfnamefont {G.-B.}\ \bibnamefont {Zhang}}, \bibinfo {author} {\bibfnamefont {X.-H.}\ \bibnamefont {Yang}}, \bibinfo {author} {\bibfnamefont {Y.-Y.}\ \bibnamefont {Ma}}, \bibinfo {author} {\bibfnamefont {Y.}~\bibnamefont {Cui}}, \bibinfo {author} {\bibfnamefont {D.-A.}\ \bibnamefont {Li}}, \bibinfo {author} {\bibfnamefont {D.-B.}\ \bibnamefont {Zou}}, \bibinfo {author} {\bibfnamefont {L.-H.}\ \bibnamefont {Du}}, \bibinfo {author} {\bibfnamefont {Z.-Q.}\ \bibnamefont {Zhao}}, \bibinfo {author} {\bibfnamefont {W.-Q.}\ \bibnamefont {Wang}},\ and\ \bibinfo {author} {\bibfnamefont {F.-Q.}\ \bibnamefont {Shao}},\ }\bibfield  {title} {\bibinfo {title} {Bubble structure evolution and electron injection controlled by optical cycles in wakefields},\ }\href {https://doi.org/10.1063/5.0156263} {\bibfield  {journal} {\bibinfo  {journal} {Physics of Plasmas}\ }\textbf {\bibinfo {volume} {30}},\ \bibinfo {pages} {073103} (\bibinfo {year} {2023})}\BibitemShut {NoStop}%
\bibitem [{\citenamefont {Zhang}\ \emph {et~al.}(2022)\citenamefont {Zhang}, \citenamefont {Chen}, \citenamefont {Zou}, \citenamefont {Zhu}, \citenamefont {Li}, \citenamefont {Yang}, \citenamefont {Liu}, \citenamefont {Yu}, \citenamefont {Ma},\ and\ \citenamefont {Sheng}}]{Zhang2022car}%
  \BibitemOpen
  \bibfield  {author} {\bibinfo {author} {\bibfnamefont {G.-B.}\ \bibnamefont {Zhang}}, \bibinfo {author} {\bibfnamefont {M.}~\bibnamefont {Chen}}, \bibinfo {author} {\bibfnamefont {D.-B.}\ \bibnamefont {Zou}}, \bibinfo {author} {\bibfnamefont {X.-Z.}\ \bibnamefont {Zhu}}, \bibinfo {author} {\bibfnamefont {B.-Y.}\ \bibnamefont {Li}}, \bibinfo {author} {\bibfnamefont {X.-H.}\ \bibnamefont {Yang}}, \bibinfo {author} {\bibfnamefont {F.}~\bibnamefont {Liu}}, \bibinfo {author} {\bibfnamefont {T.-P.}\ \bibnamefont {Yu}}, \bibinfo {author} {\bibfnamefont {Y.-Y.}\ \bibnamefont {Ma}},\ and\ \bibinfo {author} {\bibfnamefont {Z.-M.}\ \bibnamefont {Sheng}},\ }\bibfield  {title} {\bibinfo {title} {Carrier-envelope-phase-controlled acceleration of multicolored attosecond electron bunches in a millijoule-laser-driven wakefield},\ }\href {https://doi.org/10.1103/PhysRevApplied.17.024051} {\bibfield  {journal} {\bibinfo  {journal} {Phys. Rev. Appl.}\ }\textbf {\bibinfo {volume} {17}},\ \bibinfo {pages} {024051} (\bibinfo {year} {2022})}\BibitemShut {NoStop}%
\bibitem [{\citenamefont {Kim}\ \emph {et~al.}(2023)\citenamefont {Kim}, \citenamefont {Wang}, \citenamefont {Khudik},\ and\ \citenamefont {Shvets}}]{Kim2023pol}%
  \BibitemOpen
  \bibfield  {author} {\bibinfo {author} {\bibfnamefont {J.}~\bibnamefont {Kim}}, \bibinfo {author} {\bibfnamefont {T.}~\bibnamefont {Wang}}, \bibinfo {author} {\bibfnamefont {V.}~\bibnamefont {Khudik}},\ and\ \bibinfo {author} {\bibfnamefont {G.}~\bibnamefont {Shvets}},\ }\bibfield  {title} {\bibinfo {title} {Polarization and phase control of electron injection and acceleration in the plasma by a self-steepening laser pulse},\ }\href {https://doi.org/10.1088/1367-2630/acbed5} {\bibfield  {journal} {\bibinfo  {journal} {New Journal of Physics}\ }\textbf {\bibinfo {volume} {25}},\ \bibinfo {pages} {033009} (\bibinfo {year} {2023})}\BibitemShut {NoStop}%
\bibitem [{\citenamefont {Cartmell}(1990)}]{cartmell_introduction_1990}%
  \BibitemOpen
  \bibfield  {author} {\bibinfo {author} {\bibfnamefont {M.}~\bibnamefont {Cartmell}},\ }\href@noop {} {\emph {\bibinfo {title} {Introduction to linear, parametric, and nonlinear vibrations}}},\ \bibinfo {edition} {1st}\ ed.\ (\bibinfo  {publisher} {Chapman and Hall},\ \bibinfo {address} {London ; New York},\ \bibinfo {year} {1990})\BibitemShut {NoStop}%
\bibitem [{\citenamefont {Cipiccia}\ \emph {et~al.}(2011)\citenamefont {Cipiccia}, \citenamefont {Islam}, \citenamefont {Ersfeld}, \citenamefont {Shanks}, \citenamefont {Brunetti}, \citenamefont {Vieux}, \citenamefont {Yang}, \citenamefont {Issac}, \citenamefont {Wiggins}, \citenamefont {Welsh}, \citenamefont {Anania}, \citenamefont {Maneuski}, \citenamefont {Montgomery}, \citenamefont {Smith}, \citenamefont {Hoek}, \citenamefont {Hamilton}, \citenamefont {Lemos}, \citenamefont {Symes}, \citenamefont {Rajeev}, \citenamefont {Shea}, \citenamefont {Dias},\ and\ \citenamefont {Jaroszynski}}]{Cipiccia2011ga}%
  \BibitemOpen
  \bibfield  {author} {\bibinfo {author} {\bibfnamefont {S.}~\bibnamefont {Cipiccia}}, \bibinfo {author} {\bibfnamefont {M.~R.}\ \bibnamefont {Islam}}, \bibinfo {author} {\bibfnamefont {B.}~\bibnamefont {Ersfeld}}, \bibinfo {author} {\bibfnamefont {R.~P.}\ \bibnamefont {Shanks}}, \bibinfo {author} {\bibfnamefont {E.}~\bibnamefont {Brunetti}}, \bibinfo {author} {\bibfnamefont {G.}~\bibnamefont {Vieux}}, \bibinfo {author} {\bibfnamefont {X.}~\bibnamefont {Yang}}, \bibinfo {author} {\bibfnamefont {R.~C.}\ \bibnamefont {Issac}}, \bibinfo {author} {\bibfnamefont {S.~M.}\ \bibnamefont {Wiggins}}, \bibinfo {author} {\bibfnamefont {G.~H.}\ \bibnamefont {Welsh}}, \bibinfo {author} {\bibfnamefont {M.-P.}\ \bibnamefont {Anania}}, \bibinfo {author} {\bibfnamefont {D.}~\bibnamefont {Maneuski}}, \bibinfo {author} {\bibfnamefont {R.}~\bibnamefont {Montgomery}}, \bibinfo {author} {\bibfnamefont {G.}~\bibnamefont {Smith}}, \bibinfo {author} {\bibfnamefont {M.}~\bibnamefont {Hoek}}, \bibinfo {author} {\bibfnamefont {D.~J.}\ \bibnamefont {Hamilton}}, \bibinfo {author} {\bibfnamefont {N.~R.~C.}\ \bibnamefont {Lemos}}, \bibinfo {author} {\bibfnamefont {D.}~\bibnamefont {Symes}}, \bibinfo {author} {\bibfnamefont {P.~P.}\ \bibnamefont {Rajeev}}, \bibinfo {author} {\bibfnamefont {V.~O.}\ \bibnamefont {Shea}}, \bibinfo {author} {\bibfnamefont {J.~M.}\ \bibnamefont {Dias}},\ and\ \bibinfo {author} {\bibfnamefont {D.~A.}\ \bibnamefont {Jaroszynski}},\ }\bibfield  {title} {\bibinfo {title} {Gamma-rays from harmonically resonant betatron oscillations in a plasma wake},\ }\href {https://doi.org/10.1038/nphys2090} {\bibfield  {journal} {\bibinfo  {journal} {Nature Physics}\ }\textbf {\bibinfo {volume} {7}},\ \bibinfo {pages} {867} (\bibinfo {year} {2011})}\BibitemShut {NoStop}%
\bibitem [{\citenamefont {N\'emeth}\ \emph {et~al.}(2008)\citenamefont {N\'emeth}, \citenamefont {Shen}, \citenamefont {Li}, \citenamefont {Shang}, \citenamefont {Crowell}, \citenamefont {Harkay},\ and\ \citenamefont {Cary}}]{Nemeth2008laser}%
  \BibitemOpen
  \bibfield  {author} {\bibinfo {author} {\bibfnamefont {K.}~\bibnamefont {N\'emeth}}, \bibinfo {author} {\bibfnamefont {B.}~\bibnamefont {Shen}}, \bibinfo {author} {\bibfnamefont {Y.}~\bibnamefont {Li}}, \bibinfo {author} {\bibfnamefont {H.}~\bibnamefont {Shang}}, \bibinfo {author} {\bibfnamefont {R.}~\bibnamefont {Crowell}}, \bibinfo {author} {\bibfnamefont {K.~C.}\ \bibnamefont {Harkay}},\ and\ \bibinfo {author} {\bibfnamefont {J.~R.}\ \bibnamefont {Cary}},\ }\bibfield  {title} {\bibinfo {title} {Laser-driven coherent betatron oscillation in a laser-wakefield cavity},\ }\href {https://doi.org/10.1103/PhysRevLett.100.095002} {\bibfield  {journal} {\bibinfo  {journal} {Phys. Rev. Lett.}\ }\textbf {\bibinfo {volume} {100}},\ \bibinfo {pages} {095002} (\bibinfo {year} {2008})}\BibitemShut {NoStop}%
\bibitem [{\citenamefont {Curcio}\ \emph {et~al.}(2015)\citenamefont {Curcio}, \citenamefont {Giulietti}, \citenamefont {Dattoli},\ and\ \citenamefont {Ferrario}}]{Curcio2015re}%
  \BibitemOpen
  \bibfield  {author} {\bibinfo {author} {\bibfnamefont {A.}~\bibnamefont {Curcio}}, \bibinfo {author} {\bibfnamefont {D.}~\bibnamefont {Giulietti}}, \bibinfo {author} {\bibfnamefont {G.}~\bibnamefont {Dattoli}},\ and\ \bibinfo {author} {\bibfnamefont {M.}~\bibnamefont {Ferrario}},\ }\bibfield  {title} {\bibinfo {title} {Resonant interaction between laser and electrons undergoing betatron oscillations in the bubble regime},\ }\href {https://doi.org/DOI: 10.1017/S0022377815000926} {\bibfield  {journal} {\bibinfo  {journal} {Journal of Plasma Physics}\ }\textbf {\bibinfo {volume} {81}},\ \bibinfo {pages} {495810513} (\bibinfo {year} {2015})}\BibitemShut {NoStop}%
\bibitem [{\citenamefont {Horn{\'y}}\ \emph {et~al.}(2020)\citenamefont {Horn{\'y}}, \citenamefont {Kr{\r u}s}, \citenamefont {Yan},\ and\ \citenamefont {F{\"u}l{\"o}p}}]{Horny2020att}%
  \BibitemOpen
  \bibfield  {author} {\bibinfo {author} {\bibfnamefont {V.}~\bibnamefont {Horn{\'y}}}, \bibinfo {author} {\bibfnamefont {M.}~\bibnamefont {Kr{\r u}s}}, \bibinfo {author} {\bibfnamefont {W.}~\bibnamefont {Yan}},\ and\ \bibinfo {author} {\bibfnamefont {T.}~\bibnamefont {F{\"u}l{\"o}p}},\ }\bibfield  {title} {\bibinfo {title} {Attosecond betatron radiation pulse train},\ }\href {https://doi.org/10.1038/s41598-020-72053-z} {\bibfield  {journal} {\bibinfo  {journal} {Scientific Reports}\ }\textbf {\bibinfo {volume} {10}},\ \bibinfo {pages} {15074} (\bibinfo {year} {2020})}\BibitemShut {NoStop}%
\end{thebibliography}%

\end{document}